\documentclass[aps,prd,amsmath,superscriptaddress]{revtex4}

\usepackage{graphicx}
\usepackage{hyperref}

\let\l=\left
\let\r=\right
\def\be{\begin{equation}}
\def\ee{\end{equation}}
\def\bea{\begin{eqnarray}}
\def\eea{\end{eqnarray}}

\begin{document}

\title{Scalar field breathers on anti-de Sitter background}

\author{Gyula Fodor}
\affiliation{Wigner Research Centre for Physics, RMKI, 1525 Budapest 114, P.O.Box 49, Hungary}
\author{P\'eter Forg\'acs}
\affiliation{Wigner Research Centre for Physics, RMKI, 1525 Budapest 114, P.O.Box 49, Hungary}
\affiliation{LMPT CNRS-UMR 7350, Universit\'e de Tours, Parc de Grandmont, 37200 Tours, France}
\author{Philippe Grandcl\'ement}
\affiliation{LUTH, CNRS-UMR 8102, Observatoire de Paris-Meudon, place Jules Janssen, 92195 Meudon Cedex, France}

\begin{abstract}
We study spatially localized, time-periodic solutions (breathers) of scalar field theories with various self-interacting potentials on Anti-de Sitter (AdS) spacetimes in $D$ dimensions.
A detailed numerical study of spherically symmetric configurations in $D=3$ dimensions is carried out, revealing a rich and complex structure of the phase-space (bifurcations, resonances).
Scalar breather solutions form one-parameter families parametrized by their amplitude, $\varepsilon$, while their frequency, $\omega=\omega(\varepsilon)$, is a function of the amplitude.
The scalar breathers on AdS we find have a small amplitude limit, tending to the eigenfunctions
of the linear Klein-Gordon operator on AdS. Importantly most of these breathers appear to be generically stable under time evolution.
\end{abstract}

\maketitle

\today

\section{Introduction}

There has been a renewed  interest in dynamical properties of scalar fields coupled to gravity in the presence of a negative cosmological constant since the appearance of the paper \cite{bizon-11} by Bizo\'{n} and Rostworowski.
It has been shown in Ref.\ \cite{bizon-11} that generic spherically symmetric initial data evolves into a black hole state, for arbitrarily small amplitudes
of the initial data. Numerical results of Ref.\ \cite{bizon-11} indicate that energy is transferred from low to high frequency modes, reminiscent of turbulence, leading
to the instability of such space-times.
Since one expects that space-times with a negative cosmological constant, $\Lambda$, are asymptotically anti-de Sitter (AAdS),
these results are of substantial interest from the viewpoint of the AdS/CFT correspondence.
The turbulent instability, which transfers energy from low to high frequency modes and leads to black hole formation, is triggered by a resonant mode mixing.
It was then suggested that a similar mode mixing and instability already occurs in vacuum AAdS spacetimes \cite{dias-12}.
It has also  been pointed out in Ref.\ \cite{dias-12} that there is no resonant mixing if the perturbation expansion starts from a single linearized mode. The corresponding
states are periodic, so there is no black hole formation in those exceptional cases. For the pure gravity case these states are the AdS correspondents of the geons postulated by Wheeler \cite{wheeler-55,brill-64}. It was argued in another paper, that in case of negative $\Lambda$, these periodic states are nonlinearly stable \cite{dias-12b}. Time-periodic solutions of the self-gravitating massless Klein-Gordon scalar system for $\Lambda<0$ have been constructed recently in \cite{maliborski}, where their stability have been also demonstrated.

The present work is devoted to a somewhat simpler but similar system, a real scalar field, $\phi$, (both massless and massive) on a fixed anti-de Sitter background assuming spherical symmetry. The study of this system helps us to answer questions such as which part of the observed phenomena is due to the gravitational interaction, and which arise from the effectively bounded nature of the AdS spacetime due to its conformal time-like boundary.
In the fixed AdS background case there is obviously no black hole formation, but resonant mixing and turbulent instability is likely to occur in some cases. In our earlier study numerical simulation of the time-evolution of a spherically symmetric scalar field with potential $U(\phi)=\phi^2/2-\phi^4/4+\phi^6/6$ has been performed \cite{AdStimedep}. We have illustrated on a concrete example, that the energy density of initial wave-packets may increase by 5 orders of magnitude after a few hundred reflections from the time-like boundary of AdS space-time. Such a drastic increase of the energy density seems to strongly depend on the potential, in particular for the standard $\phi^4$ potential, $U(\phi)=\phi^2(\phi-2)^2/8$, and also for $U(\phi)=\phi^4$, no such dramatic energy density increase has been seen.

In this paper we carry out a detailed study of spherically symmetric, spatially localized, time-periodic scalar field solutions on a fixed AdS background, when the nonlinearity is provided by the scalar field interaction potential $U(\phi)$.
It is well known that on AdS space-times a free scalar field theory (linear Klein-Gordon theory, i.e.\ $U(\phi)=m^2\phi^2/2$) even for $m=0$ admits
regular, spatially localized, time-periodic breather-type solutions \cite{avis}. In fact these solutions are known analytically, and
they form a discretely infinite family indexed by the nodes of the scalar field. As a matter of fact, when a real scalar field is coupled to gravitation on AAdS space-times the wave equation continues
to admit a discretely infinite family of breathers even when the scalar field is massless ($m=0$) \cite{isibashiwald,maliborski}.

There are breather solutions on a fixed Ads background for general $U(\phi)$ potentials which have a minimum. 
In this paper the localized oscillating states on AdS are called breathers, because they are truly localized and oscillate exactly periodically. On flat background the only such state is the well known sine-Gordon breather, which exists only for the sine-Gordon potential $U(\phi)=1-\cos\phi$, and even for that potential, only in one spatial dimension. Interestingly, there are very long living localized oscillating states even on flat Minkowski background, which are called oscillons\cite{BogMak2,Gleiser,CopelGM95}, for other potentials or other dimensions, but they are not exactly periodic. They emit energy very slowly, accompanied by a slow change of their amplitude and frequency  \cite{Honda,oscillons,FFHL2}. It has been shown that the energy emission rate of small-amplitude oscillons is exponentially small in terms of a parameter corresponding to the central amplitude \cite{SK,FFHM,moredim}. Flat background oscillons only exist for massive fields, since the frequency of small amplitude oscillons is
determined by the scalar field mass. Furthermore, the size of small-amplitude oscillons grows without limit, inversely proportionally to their decreasing amplitude. On the other hand, breathers on AdS background also exist for zero mass scalars, since their size in the small-amplitude limit tends to a finite value determined by the cosmological constant. Moreover, the frequency of AdS breathers formed by zero mass scalars approaches the number of spatial dimensions in the theory.

Massive scalar fields coupled to Einstein's gravity also can form long living oscillating localized objects, which in the $\Lambda=0$ case are called oscillatons in the literature\cite{Seidel1,Seidel2,page04}. Oscillatons are not exactly periodic, but they are extremely long-living, the mass loss rate of oscillatons is exponentially small in terms of an amplitude parameter\cite{oscillaton10,oscillatons}. Unlike flat background oscillons, self gravitating oscillatons can be formed from a massive Klein-Gordon field, since gravity provides the necessary nonlinearity. Similar objects to self-gravitating oscillatons, but with AdS asymptotics, are also expected to exist for $\Lambda<0$, in which case they should be exactly periodic, so they might be called breathers as well. The time-periodic solutions formed by self-gravitating massless Klein-Gordon fields with $\Lambda<0$ have been studied recently in \cite{maliborski}.

In the small amplitude limit the AdS breathers we construct are nonlinear deformation of of the linear Klein-Gordon (KG) breathers, existing only for a discrete set of frequencies. The nonlinearity induced by the self-interaction potential leads to one parameter families of solutions parametrized by their amplitudes, with their frequencies
$\omega=\omega(\varepsilon)$, reducing in the limit $\varepsilon\to0$ to the corresponding value of that of the KG breather. In the small amplitude limit 
our perturbative construction permits to deduce analytically the frequency-amplitude relation to leading order
in arbitrary dimensions and potentials. These perturbative results are found to be in excellent agreement with
high-precision numerical computations.  
Medium amplitude states can be described by perturbational expansion methods, but large amplitude breathers can only be studied by direct numerical analysis.

The obtained periodic solutions for the scalar field on fixed AdS background necessarily contain some numerical round-off errors. Hence using them as initial data for a numerical time-evolution code provide information about their stability. This type of analysis strongly indicates that up to a certain amplitude the breather solutions are nonlinearly stable. Remarkably not only the basic (nodeless) breathers are stable, but even those with several nodes. One can expect that it is easier to find stable periodic solutions on Ads background than on Minkowski background, since in this case energy cannot dissipate to infinity. The natural boundary condition at AdS infinity, which is generally used for a minimally coupled scalar fields, implies that energy dissipated towards infinity gets back to the central region in a finite time again. Among the larger amplitude periodic states there are unstable configurations. These do not decay completely, but keep on oscillating, however with an amplitude which is generally a complicated function of
time. Their evolution does not show sign of turbulent instability either, energy is not transferred continuously to high frequency short wavelength modes.

An interesting feature of the numerically constructed breather solutions is that if we plot the central amplitude as a function of the frequency then thin resonant peaks appear on the plot at certain frequencies. Fourier decomposing the time dependence of the field, it appears that near these resonances one of the higher Fourier modes starts to grow. The spatial profile of these growing modes turns out to be very similar to that of a Klein-Gordon breather with several nodes. Furthermore the frequency of the corresponding Klein-Gordon breather is relatively close to the frequency of the resonant peak. Nonlinear effects are responsible for a shift in the frequency, and further studies would be worthwhile in order to understand the resonant peaks better. We note that similar resonant frequencies already occur in the much simpler system, where a single scalar field evolves on a one dimensional flat torus \cite{GenMastroProc}. Numerical evolution shows that generally there are unstable periodic states close to these
resonances.

The paper is organized as follows. In Sec.~\ref{sec-scalars} we present the used coordinate systems and the equations describing the scalar fields. Section \ref{kleingordon} is devoted to the detailed study of general time-periodic solutions of the linear differential equations describing massive or massless Klein-Gordon fields on AdS background. It turns out that for solutions with finite energy the scalar field necessarily tends to zero at large radiuses, so the finite energy solutions are the same as the localized solutions in this system. Localized solutions with a regular center only exist for a discrete set of frequencies. However, all periodic solutions will be necessary later, when we will have to solve inhomogeneous differential equations at the small-amplitude expansion in the nonlinear case. In Sec.~\ref{spectral} the high precision spectral method is described, which is used to get localized time-periodic breather solutions in this paper. As a test for the code the periodic Klein-Gordon breathers
are constructed, without a priori knowledge of the allowable frequencies. Theses solutions will be used later in the nonlinear case as starting points for the iterative process in the spectral code. Section \ref{selfinter} starts with the presentation of the considered scalar field self-interaction potentials, and continues with the introduction of a rescaled scalar field variable which takes into account the asymptotic falloff properties. In Subsection \ref{persolsec} the spectral code is applied to the case of a symmetric double well potential, and numerical results are presented for three different values of the cosmological constant. In each cases the Fourier modes of the scalar field are shown as functions of the oscillation frequency. The results of the spectral code are used as initial data for a fourth-order time evolution code in Subsection \ref{timeevol}. This allows us to investigate the stability properties of the breather solutions. It turns out that there are unstable solutions for high
amplitudes, and that configurations close to some resonant frequencies may be also unstable. In Sec.~\ref{smallamplsec} a small-amplitude expansion procedure is presented for general $U(\phi)$ potentials. This approach is based on the property that in the small-amplitude limit the periodic solutions become similar to the Klein-Gordon breathers presented in Sec.~\ref{kleingordon}. The comparison with the precise numerical results of Sec.~\ref{spectral} shows that the expansion procedure is valid up to quite large amplitudes. In Sec.~\ref{smalksec} a different approach to build up a small-amplitude expansion procedure is presented, which is useful when the cosmological constant is small. This approach is a generalization of the procedure which has been used earlier for the expansion of flat background oscillon states.

\section{Scalars in anti-de Sitter spacetime} \label{sec-scalars}

We consider $1+D$ dimensional anti-de Sitter spacetime, which is a maximally symmetric space with scalar curvature $R=2\frac{D+1}{D-1}\Lambda$, where $\Lambda$ is a negative cosmological constant. Choosing any geodesic observer, its world line can be the center of a global static coordinate system. One such coordinate system for AdS is based on Schwarzschild area coordinates, and the line element is
\begin{equation}
 ds^2=-(1+k^2r^2)dt^2+\frac{dr^2}{1+k^2r^2}+r^2d\Omega_{D-1}^2 \ , \label{schcoord}
\end{equation}
where $k$ is a constant related to $\Lambda$ and the coordinates are $x^a=(t,r,\theta_1\ldots\theta_{D-1})$. This metric satisfies the vacuum Einstein equations $G_{ab}+\Lambda g_{ab}=0$ with cosmological constant
\begin{equation}
 \Lambda=-\frac{1}{2}D(D-1)k^2 \ .
\end{equation}
Another very useful coordinate system can be obtained by setting
\begin{equation}
 r=\frac{1}{k}\tan x \ , \qquad t=\frac{1}{k}\tau \ .
\end{equation}
This way we obtain the form of the anti-de Sitter metric which is conformal to half of the Einstein static universe,
\begin{equation}
 ds^2=\frac{1}{k^2\cos^2x}\left(-d\tau^2+dx^2+\sin^2x\, d\Omega_{D-1}^2\right) . \label{confcord}
\end{equation}
The origin is at $x=0$, infinity is at $x=\pi/2$, and the time coordinate $\tau$ runs from minus to plus infinity. When AdS is embedded in $2+D$ dimensional Minkowski spacetime, it becomes time-periodic in $\tau$ with period $2\pi$.

The action of a minimally coupled real scalar field on a curved background is
\begin{equation}
 A=\int dtd^D x \mathcal{L} \ , \qquad
 \mathcal{L}=-\sqrt{-g}\left[\frac{1}{2}g^{ab}\phi_{,a}\phi_{,b}+U(\phi)\right] ,
\end{equation}
where $U(\phi)$ is a scalar potential. The field equation is
\begin{equation}
g^{ab}\phi_{;ab}=U'(\phi) \ , \label{waveeq1}
\end{equation}
and the stress-energy tensor can be written as
\begin{equation}
 T_{ab}=-\frac{2}{\sqrt{-g}}\,\frac{\delta A}{\delta g^{ab}}=\phi_{,a}\phi_{,b}-g_{ab}\left[\frac{1}{2}g^{cd}\phi_{,c}\phi_{,d}+U(\phi)\right]  . \label{streneq}
\end{equation}

We note that since for AdS the curvature is constant, conformally invariant scalar fields can also be described by equation \eqref{waveeq1} after making the substitution
\begin{equation}
U(\phi) \longrightarrow U(\phi)-\frac{1}{8}(D^2-1)k^2\phi^2 \ .
\end{equation}
This means that coefficient of the $m^2\phi^2/2$ term in $U(\phi)$ is shifted by a negative number, possibly making the square of the scalar field mass $m$ negative. In this paper we only consider minimally coupled scalar fields in the following, so we assume $m^2\geq0$.

The field equation for a spherically symmetric scalar field using the Schwarzschild coordinate system is
\begin{equation}
 -\frac{1}{1+k^2r^2}\phi_{,tt}+(1+k^2r^2)\phi_{,rr}+\frac{1}{r}\left[D-1+(D+1)k^2r^2\right]\phi_{,r}=U'(\phi) \  .  \label{fieldeq}
\end{equation}
Using the timelike Killing vector $k^a=\delta^a_0$ we can define the conserved current vector $J^a=T^{ab}k_b$. The unit normal vector to the constant $t$ hypersurfaces is $t^a=\delta^a_t/\sqrt{1+k^2r^2}$, and the energy density of the configuration, $\epsilon=t^aJ_a$, can be calculated from \eqref{streneq} as
\begin{equation}
\epsilon=\sqrt{1+k^2r^2}
\left[\frac{\left(\phi_{,t}\right)^2}{2(1+k^2r^2)}+\frac{1}{2}(1+k^2r^2)\left(\phi_{,r}\right)^2+U(\phi)\right]  . \label{schvarepseq2}
\end{equation}
The energy inside a sphere given by $r=r_1$ is then
\begin{equation}
E=\int_0^{r_1} \mathcal{E}\,dr  \ , \label{einteq}
\end{equation}
where
\begin{equation}
\mathcal{E}=\frac{2\pi^{D/2}}{\Gamma(D/2)}\,\frac{r^{D-1}}{\sqrt{1+k^2r^2}}\,\epsilon \ .
\end{equation}

The field equation for the scalar field using the conformal coordinate system \eqref{confcord} is
\begin{equation}
 -\phi_{,\tau\tau}+\phi_{,xx}+\frac{D-1}{\sin x\cos x}\,\phi_{,x}=
\frac{1}{k^2\cos^2 x}U'(\phi) \  .  \label{cfieldeq2}
\end{equation}
The timelike Killing vector field, which has unit norm at the center is now $k^a=k\delta^a_\tau$, and we define the conserved current vector by $J^a=T^{ab}k_b$. The normal vector to the constant $\tau$ surfaces is $t^a=k\cos x\delta^a_\tau$, and the energy density is
\begin{equation}
\epsilon=t^aJ_a=k^2\cos x\left[\frac{1}{2}\left(\phi_{,\tau}\right)^2
+\frac{1}{2}\left(\phi_{,x}\right)^2
+\frac{1}{k^2\cos^2x}U(\phi)\right]  \  . \label{confvarepseq2}
\end{equation}
The energy inside a sphere $x=x_1$ is then
\begin{equation}
E=\int_0^{x_1} \mathcal{E}\, dx  \ , \label{einteqconf}
\end{equation}
where
\begin{equation}
\mathcal{E}=\frac{2\pi^{D/2}}{\Gamma(D/2)}\,\frac{1}{k\cos x}\left(\frac{\tan x}{k}\right)^{D-1}\epsilon \ .
\end{equation}

\section{Klein-Gordon breathers on AdS}
\label{kleingordon}

We study Klein-Gordon fields with potential $U(\phi)=m^2\phi^2/2$ in this section in more detail because there are periodic localized exact solutions on AdS background in this case. Furthermore, the asymptotic behavior at large distances for general potentials can be well approximated by the behavior of the same mass Klein-Gordon field. We look for periodic solutions in the form \begin{equation}
\phi=p\cos\left(\frac{\omega}{k}\tau\right) \ , \label{oneharm}
\end{equation}
where the function $p$ depends only on $x$. From \eqref{cfieldeq2} follows that $p$ satisfies the ordinary differential equation
\begin{equation}
p_{,xx}+\frac{D-1}{\sin x\cos x}\,p_{,x}
=\frac{m^2}{k^2\cos^2 x}p-\frac{\omega^2}{k^2}p \ . \label{pfeq}
\end{equation}
We note that $t=\tau/k$ is the physical time, measuring the proper time of the geodesic observer at the origin, so $\omega$ is the physical (angular) frequency of the oscillating field.

One of the two independent solutions of \eqref{pfeq} near the center $x=0$ tends to a constant, while the other solution diverges as $x^{2-D}$ for $D\geq 3$, or logarithmically for $D=2$. Hence for $D\geq 2$ the assumption that $p$ remains finite near $x=0$ ensures that the scalar field $\phi$ is regular at the center and satisfies the field equation there. Also follows that in the series expansion of $p$ at $x=0$ there are only even powered terms of $x$, so $p$ has a mirror symmetry there. We can impose this mirror symmetry as a boundary condition at $x=0$ for the $D=1$ spatial dimensional case.

Let us introduce the coordinate $y$, describing the coordinate distance from infinity, by $y=\pi/2-x$. Expanding around $y=0$ it turns out that there are two independent solutions of the homogeneous linear equation \eqref{pfeq},
\begin{equation}
p^\pm=y^{\lambda_\pm}\left(1+p_2^\pm y^2+p_4^\pm y^4+\ldots\right) ,
\end{equation}
where
\begin{equation}
\lambda_\pm=\frac{D}{2}\pm\sqrt{\frac{D^2}{4}+\frac{m^2}{k^2}} \ ,
\label{lambdapm}
\end{equation}
and $p_i^\pm$ are coefficients depending on $D$, $k$, $m$ and $\omega$. Note that since $\lambda_\pm$ does not contain $\omega$, the leading order asymptotic behavior is frequency independent. Since in this paper we only consider minimally coupled scalar fields, necessarily $m^2\geq0$, and consequently $\lambda_-\leq0$ and  $\lambda_+>0$. The solution $p^+$ is localized in the sense that it tends to zero at infinity, but $p^-$ is not. For zero mass fields $\lambda_+=D$ and $\lambda_-=0$.

According to \eqref{einteqconf}, the total energy $E$ of a configuration can be calculated by integrating the function $\mathcal{E}$ in the $[0,\pi/2]$ interval. If the field behaves as $\phi\approx cy^\lambda\cos(\omega\tau/k)$ near infinity $y=0$, then from \eqref{confvarepseq2} to leading order the function $\mathcal{E}$ behaves like
\begin{equation}
\mathcal{E}\approx \frac{c^2\pi^{D/2}y^{2\lambda-D-1}}{\Gamma(D/2)k^D}
\left(m^2+k^2\lambda^2\right)\cos^2\left(\frac{\omega\tau}{k}\right) . \label{caleasy}
\end{equation}
This has a finite integral if and only if $\lambda>D/2$. Consequently, according to \eqref{lambdapm}, a localized solution, which tend to zero as $y^{\lambda_+}$ at infinity, always has finite energy. On the other hand, if a solution behaves as $y^{\lambda_-}$, it necessarily has infinite energy. Because of this, we are interested in solutions of \eqref{pfeq} for which $p$ tends to zero as $(\pi/2-x)^{\lambda_+}$ at the boundary $x=\pi/2$.

In order to be able to write periodic Klein-Gordon solutions in terms of hypergeometric functions it is useful to introduce a new radial coordinate $z$ by
\begin{equation}
z=\cos^2 x=\frac{1}{1+k^2r^2} \ . \label{zcoord}
\end{equation}
In terms of this coordinate $z=0$ corresponds to infinity and $z=1$ to the center. Close to infinity this new coordinate is related to the previously defined $y=\pi/2-x$ by $z\approx y^2$, and hence the asymptotic falloff of $p$ at infinity is $z^{\lambda_+/2}$. Using the radial coordinate $z$, equation \eqref{pfeq} describing periodic Klein-Gordon configurations becomes
\begin{equation}
z^2(1-z)p''+z\left(1-z-\frac{D}{2}\right)p'+\frac{1}{4k^2}\left(\omega^2z-m^2\right)p=0 \  ,
\label{pzeq}
\end{equation}
where the prime denotes differentiation with respect to $z$. The boundary condition we require in order to have localized regular breathers is that $p$ tends to a constant at $z=1$ and approaches zero at $z=0$ as $z^{\lambda_+/2}$. In terms of the coordinate $z$ there are both even and odd power terms in the Taylor expansion of $p$ around $z=1$, so there is no mirror symmetry now.

We would like to transform \eqref{pzeq} into the form of a hypergeometric differential equation\cite{dlmf}
\begin{equation}
z(1-z)q''+\left[c-(a+b+1)z\right]q'-abq=0 \ .  \label{hipgeoeq}
\end{equation}
In order to achieve this, we define a rescaled field variable $q$ by $p=z^\alpha q$, substitute into \eqref{pzeq}, and look for the condition of the vanishing of the $z^\alpha q$ term, in order to ensure that the equation should have an overall $z^{\alpha+1}$ factor. This gives that $\alpha=\lambda_\pm/2$, where $\lambda_\pm$ are the constants defined earlier in \eqref{lambdapm}. Since it can be shown that both choices give the same set of solutions, we chose the plus sign, and define $q$ by
\begin{equation}
p=z^{\lambda_+/2} q \ .
\end{equation}
We have seen that localized solutions have exactly this asymptotic behavior, consequently the rescaled field variable $q$ tends to a finite constant at $z=0$. Substituting into \eqref{pzeq} we obtain
\begin{equation}
D_z q-\frac{1}{4}\left(\lambda_+^2-\frac{\omega^2}{k^2}\right)q=0 \ , \label{dzqeq}
\end{equation}
where we define the differential operator $D_z$ by
\begin{equation}
D_z=z(1-z)\partial_{z}^2+\left[(1-z)(\lambda_++1)-\frac{D}{2}\right]\partial_{z} \ . \label{dzdef}
\end{equation}
In order to obtain a localized breather solution with a regular center we have to impose the boundary condition that $q$ tends to a finite constant both at $z=0$ and $z=1$. At both ends the breather solution $q$ can be expanded in Taylor series, but the expansion will contain both even and odd powers, so there will be no mirror symmetry.

Equation \eqref{dzqeq} is just the intended hypergeometric equation, and the constants in the canonical form \eqref{hipgeoeq} are
\begin{equation}
a=\frac{1}{2}\left(\lambda_++\frac{\omega}{k}\right) \ , \quad
b=\frac{1}{2}\left(\lambda_+-\frac{\omega}{k}\right) \ , \quad
c=\lambda_+-\frac{D}{2}+1 \ .  \label{abceq}
\end{equation}
A pair of fundamental solutions of the hypergeometric differential equation \eqref{hipgeoeq}, which are numerically satisfactory near infinity $z=0$ are (see Sec.~15.10 of \cite{dlmf})
\begin{equation}
w_1=\,_2F_1(a,b;c;z) \quad , \quad
w_2=z^{1-c}\,_2F_1(a-c+1,b-c+1;2-c;z) \ . \label{homsolz0}
\end{equation}
Fundamental solutions around the center $z=1$ are
\begin{equation}
w_3=\,_2F_1(a,b;a+b-c+1;1-z) \quad , \quad
w_4=(1-z)^{c-a-b}\ _2F_1(c-a,c-b;c-a-b+1;1-z) \ . \label{homsolz1}
\end{equation}
Since the hypergeometric function $_2F_1(a,b;c;z)$ smoothly tends to $1$ at $z=0$, the solution $w_1$ is always regular at infinity $z=0$, while $w_3$ is always regular at the center. Since $1-c=D/2-\lambda_+<0$, the solution $w_2$, if it is well defined by \eqref{homsolz0}, is necessarily singular at $z=0$. Similarly, since $c-a-b=1-D/2<0$ the solution $w_4$ is singular at $z=1$ for $D\geq3$. The hypergeometric functions in the solutions are not defined when their third argument is a non-positive integer. This cannot happen for $w_1$, since always $c>0$, and cannot happen for $w_3$ either, since $a+b-c+1=D/2$. Consequently, the periodic Klein-Gordon field solutions that oscillate with frequency $\omega$ and have a regular center are determined by $w_3$, and can be written as \cite{avis}
\begin{equation}
\phi=A\cos\left(\frac{\omega}{k}\tau\right)(\cos x)^{\lambda_+}
\,_2F_1\left(
\frac{1}{2}\left(\lambda_++\frac{\omega}{k}\right),
\frac{1}{2}\left(\lambda_+-\frac{\omega}{k}\right);
\frac{D}{2};\sin^2 x\right) ,  \label{phipergen}
\end{equation}
where $A$ is an arbitrary constant. All other solutions have a singularity at the center of symmetry. However, for general $\omega$ this solution diverges at infinity. In order to study the behavior of the periodic solution \eqref{phipergen} at infinity $z=0$, we can apply the identity\cite{dlmf}
\begin{equation}
w_3=\frac{\Gamma(1-c)\Gamma(a+b-c+1)}{\Gamma(a-c+1)\Gamma(b-c+1)}w_1
+\frac{\Gamma(c-1)\Gamma(a+b-c+1)}{\Gamma(a)\Gamma(b)}w_2 \ . \label{w3trans}
\end{equation}
Since $w_2$ is always singular at $z=0$, the solution $w_3$ is also singular unless the coefficient of $w_2$ in \eqref{w3trans} is zero. Taking into account \eqref{abceq}, this can only happen if $\Gamma(b)$ is singular, which happens if $b$ is a non-positive integer. We obtain that the periodic scalar field solution \eqref{phipergen} is regular both at the center and at infinity if and only if the frequency $\omega$ takes the discrete values
\begin{equation}
\omega=k(\lambda_++2n) ,  \label{lambdapcond}
\end{equation}
where $n\geq0$ integer. In this case the series defining the hypergeometric function becomes a finite sum, and $w_3$ becomes an $n$-th order polynomial in $1-z$. It follows that a minimally coupled Klein-Gordon scalar field on AdS background can form regular localized breather solutions if and only if \eqref{lambdapcond} holds for a nonnegative integer $n$, and the solutions can be written as\cite{avis}
\begin{equation}
\phi^{(n)}_{KG}=A_n\frac{n!}{(D/2)_n}\cos\left[(\lambda_++2n)\tau\right](\cos x)^{\lambda_+}P_{n}^{(D/2-1,\lambda_+-D/2)}\left(\cos(2x)\right) ,  \label{kgbreather}
\end{equation}
where $P$ denotes the Jacobi polynomial, $(\alpha)_n=\alpha(\alpha+1)\ldots(\alpha+n-1)$ denotes the Pochhammer symbol, with $(\alpha)_0=1$, and $A_n$ are arbitrary constants giving the value of $\phi^{(n)}_{KG}$ at the center $x=0$.

The transformation rule \eqref{w3trans} cannot be applied if the second solution $w_2$ is not defined by \eqref{homsolz0}, which happens if $2-c=1+D/2-\lambda_+$ is a non-positive integer. For example, this is the case when $m=0$ and $D$ is even, since for zero mass fields $\lambda_+=D$. It is possible to prove that the condition \eqref{lambdapcond} has to hold even in these cases in order to have localized breathers\cite{balasub,isibashiwald}, and the form of the breathers is also given by \eqref{kgbreather}.

\begin{figure}[!hbtp]
\includegraphics[width=10cm]{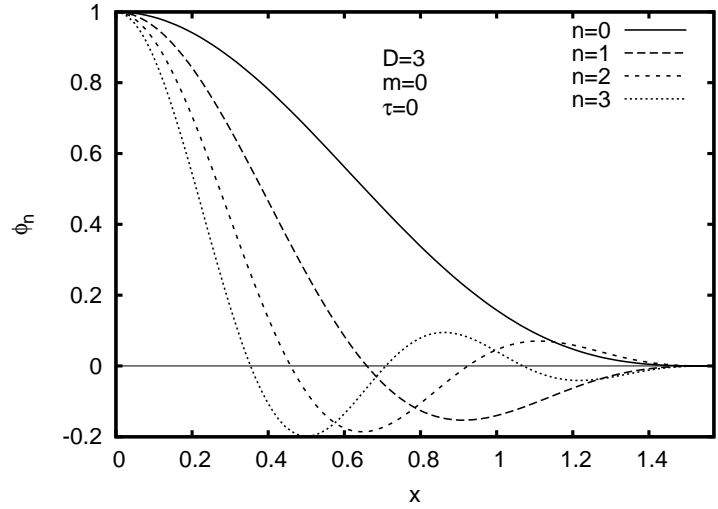}
\caption{Breather configurations with various nodes for massless scalar field. \label{breatherfig}}
\end{figure}

\begin{figure}[!hbtp]
\includegraphics[width=10cm]{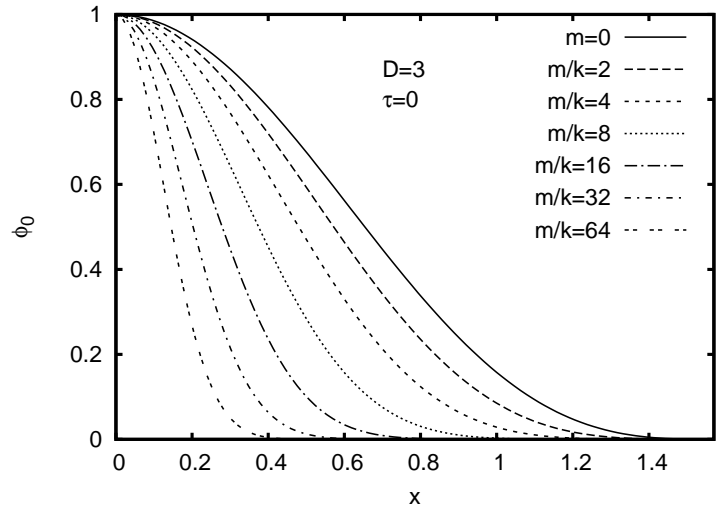}
\caption{Nodeless breather configurations for various $m/k$ values. \label{brmchfig}}
\end{figure}

The breather $\phi^{(n)}_{KG}$ has $n$ nodes. The first few solutions are
\begin{align}
 \phi^{(0)}_{KG}&=A_0\cos\left(\lambda_+\tau\right)(\cos x)^{\lambda_+} \ ,\\
 \phi^{(1)}_{KG}&=A_1\cos\left[(\lambda_++2)\tau\right](\cos x)^{\lambda_+}\left[1-\frac{2(\lambda_++1)}{D}\sin^2 x\right]  ,\\
 \phi^{(2)}_{KG}&=A_2\cos\left[(\lambda_++4)\tau\right](\cos x)^{\lambda_+}\left[1
 -\frac{4(\lambda_++2)}{D}\sin^2 x
 +\frac{4(\lambda_++2)(\lambda_++3)}{D(D+2)}\sin^4 x\right] .
\end{align}
For a massless field in $D=3$ spatial dimensions the first four curves of $\phi^{(n)}_{KG}$ at $\tau=0$ are shown on Fig.~\ref{breatherfig}.
For massive fields the breathers become more and more compact as $m$ is increases, as it is shown on Fig.~\ref{brmchfig}.

For zero mass scalar fields $\lambda_+=D$, and for odd spatial dimensions it may be possible to simplify the expression \eqref{kgbreather} for the $n$ node breather. For example, for $D=1$ the zero scalar mass breathers are simply
\begin{equation}
\phi^{(n)}_{KG}=A_n\cos[(1+2n)\tau]\cos[(1+2n)x] \ ,
\end{equation}
which is not surprising, since the field equation \eqref{cfieldeq2} is then simply $-\phi_{,\tau\tau}+\phi_{,xx}=0$. But even then, the background metric \eqref{confcord} is nontrivial. For the physically more important $D=3$ case
\begin{equation}
\phi^{(n)}_{KG}=\frac{A_n\cos[(3+2n)\tau]}{4\sin x}\left(\frac{\sin[2(1+n)x]}{1+n}+
\frac{\sin[2(2+n)x]}{2+n}\right)
\end{equation}
for a zero mass Klein-Gordon field.

Both for massive and massless fields the asymptotic behavior at infinity is given by $(\cos x)^{\lambda_+}\sim y^{\lambda_+}$. All regular finite energy solutions of the linear Klein-Gordon equation on AdS can be expanded as sums of $\phi^{(n)}_{KG}$ with appropriate phase shifts in the time coordinate. The breather configurations are stable in the sense that if an initial data is chosen close to them, then the time evolution remains close to them as well. For zero mass fields $\lambda_+=D$, and $\omega/k=D+2n$. For large scalar mass or for small $k$, i.e.~when $m/k\gg D$ we have $\lambda_+\approx m/k$, and $\omega\approx m+2nk$.

The energy of the Klein-Gordon breather solution \eqref{kgbreather} can be calculated from \eqref{einteqconf},
\begin{equation}
E_n=\frac{A_n^2n!\pi^{D/2}\Gamma\left(\lambda_+-\frac{D}{2}\right)}{4k^D(D/2)_n\Gamma(\lambda_++n)}
\left(\lambda_+-\frac{D}{2}+1\right)_n
\left\{2m^2+k^2\left[D\lambda_++4n\left(\lambda_+-\frac{D}{2}\right)\right]\right\} .
\end{equation}

\section{Spectral solver}\label{spectral}

\subsection{General setting}
The numerical library Kadath \cite{kadath, kadathpaper} is employed throughout this paper to compute the structure of the breathers. It is used for all the numerical parts that do not involve explicit time evolution of the field. The Kadath library enables the use of spectral methods for a large class of problems arising in physics.

The setting is very similar to the one already used in our work on oscillatons \cite{oscillatons} with a use of a two-dimensional space. One of the dimensions is the radial coordinate and the other one is the time. In terms of the radial coordinate, the computational space is decomposed in various numerical domains. In each domain, the numerical coordinate $r^\star$ relates to the physical one by an affine-law so that $r^\star$ varies in the right interval to perform the spectral expansion (i.e. in $\l[-1, 1\r]$ for all domains except the one containing the origin for which it lies in $\l[0, 1\r]$ for parity reasons). The physical radius itself can span $\l[0, \infty\r[$ if one uses the Schwarzschild area coordinates (\ref{schcoord}) or $\l[0, \pi/2\r[$ for the compactified coordinates (\ref{confcord}). A single domain is used in time for which the coordinate time $t^\star$ relates to the physical one by $t^\star = \omega t$, where $\omega$ is the frequency of the configuration. For parity reasons, $t^\star$ covers only the interval $\l[0, \pi\r]$.

In each domain, the field is expanded onto spectral basis with respect to the coordinates $\l(r^\star, t^\star\r)$. For the radial coordinate, one uses the Chebyshev polynomials $T_i \l(r^\star\r)$ (except in the domain containing the origin where only even Chebyshev polynomials are used for regularity reasons). Given the parity of the configurations, one uses only cosines for the time variable and so expand the field into functions like $\cos\l(j t^\star\r)$.

Partial differential equations on functions are transformed into algebraic equations on the spectral coefficients by making use of the so-called weighted residual method. The obtained non-linear system is then solved by iteration by means  of a Newton-Raphson iteration. We refer the reader to \cite{living, kadathpaper} for more details about spectral methods and Kadath.

\subsection{Equations}
When using the non-compact radial coordinate, numerical computations are performed using a variation of Eq. (\ref{fieldeq}). The only difference is a rescaling of $r$ and $t$ with respect to $k$ so that the equation takes the form:

\begin{equation}
 -\frac{1}{1+r^2}\phi_{,tt}+(1+r^2)\phi_{,rr}+\frac{1}{r}\left[2r^2+(D-1)(1+r^2)\right]\phi_{,r}=\frac{U'(\phi)} {k^2}\  .  \label{fieldeqnum}
\end{equation}

Near the origin the term in $\phi_{,r}/{r}$ is computed in the coefficient space to ensure regularity ($\phi$ being even near the origin, its radial derivative can always be divided by zero). In the domain containing the spatial infinity, we divide Eq. (\ref{fieldeqnum}) by $(1+r^2)$ to prevent the appearance of infinite quantities. The equation is supplemented with the boundary condition that $\phi$ vanishes at $r=\infty$.

When a compact coordinate is used, we implement the field equation (\ref{cfieldeq2}). In order to avoid divergence at $x=\pi/2$ the equation is multiplied by $\cos^2 x$ which gives
\begin{equation}
\cos^2x \l(-\phi_{,\tau\tau}+\phi_{,xx}\r)+\l(D-1\r)\frac{\cos x}{\sin x}\,\phi_{,x}=
\frac{1}{k^2}U'(\phi) \  .  \label{cfieldeq2num}
\end{equation}
In the central region, the term in $\phi_{,x} / \sin x$ is computed in the coefficient space and is always regular due to the parity of $\phi$.

Let us mention than a third version of the field equation can be used in the case of a self-interacting field (see Sec. \ref{selfinter}).

\subsection{Getting the Klein-Gordon breathers}

In this section we consider only massive Klein-Gordon fields with $m=1$, meaning we take $U\l(\phi\r) = \phi^2 /2$. Despite their apparent simplicity, the numerical solution of Eqs (\ref{fieldeqnum}) or (\ref{cfieldeq2num}) requires special care and a dedicated numerical treatment. In a usual Kadath setting, one starts from an initial guess as good as possible and then solve the system of equations by iteration. In this particular case, we know that there exist solutions only for a discrete set of values for $\omega$ given by Eq.~(\ref{lambdapcond}). However, one wishes to solve the problem numerically without making use of this value beforehand, so that one needs to find a numerical method to find those discrete values. Even when $\omega$ is set to an appropriate value, the equation being linear, it admits a whole family of solution. One needs to find a way to select one particular solution (for instance the one for which the field is $1$ at the origin) and in the same way avoid the trivial solution of a field that is identically zero. The following method takes care of all those difficulties.

The idea is to split the space into two regions : the inner part for $r<r_{\rm lim}$ and the outer part for $r>r_{\rm lim}$, and to solve the equation independently in the two regions. The choice of $r_{\rm lim}$ has no importance. To close both systems of equations one must provide a boundary condition at $r_{\rm lim}$, that is we impose that $\phi\l(r_{\rm lim}\r) = V_{\rm lim}$. The equation being linear, the value $V_{\rm lim}$ is not expected to matter, as long as it is not zero. Doing so, one gets, in each region, a solution of the field equation that can not be identically zero, due to the boundary condition that is enforced.

\begin{figure}[!hbtp]
\includegraphics[width=10cm]{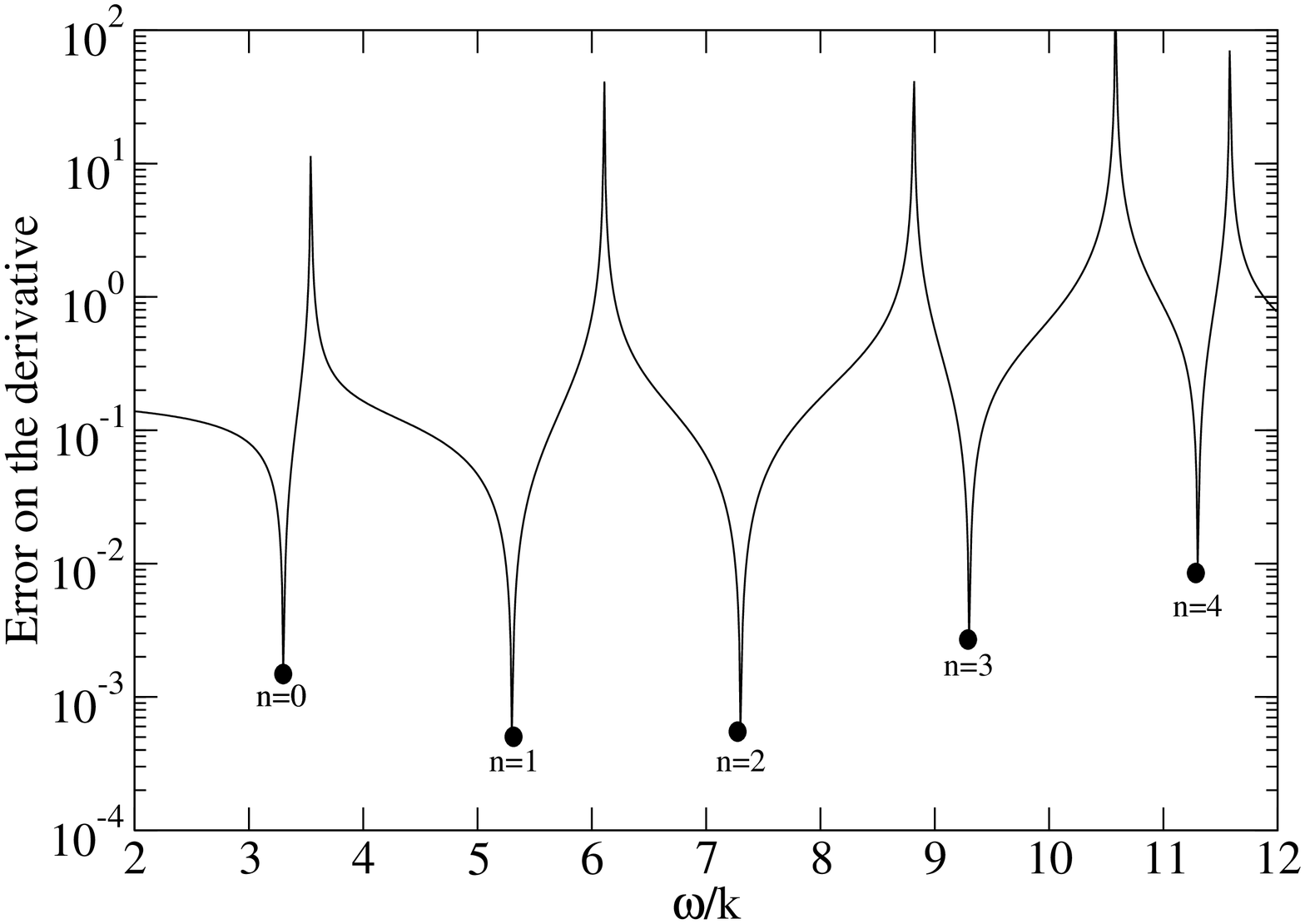}
\caption{Error on the derivative, measured by $\l|\partial_r \phi\l(r=r_{\rm lim}^-\r) - \partial_r \phi\l(r=r_{\rm lim}^+\r)\r|$ , as a function of $\omega/k$. The figure is done for $m=1$ and $k=1$. The various filled circles indicate the values for which a true solution found, labeled with respect to the number of nodes. \label{figomega}}
\end{figure}

However, at the radius between the two regions, the solution is, in general, not $\mathcal{C}^\infty$. Indeed, given the fact that the equation is of second order and that one has only ensured the continuity of the solution itself (via the boundary condition), the radial derivative is generally not continuous. It is precisely this additional condition that constraints the value of $\omega$. If one computes the discontinuity in the derivative of the solution as a function of $\omega$, as is done in Fig.~\ref{figomega}, one finds that it does vanish only for a discrete set of values. Those correspond to the frequencies that do lead to a true solution of the system and they are analytically given by Eq.~(\ref{lambdapcond}). The different zeros correspond to increasing values of $n$. The various admissible values of $\omega$ are numerically found to the desired precision by using the secant method to find the location of the zeros of the error. The technique can be used also with the compactified coordinate system. Fig.~\ref{figomega} shows the difference in the radial derivative on the two sides of $r_{\rm lim}$ as a function of $\omega/k$, for $k=1$. The various location of the zeros are given by the black circles and the labels give the associated number of nodes of the solution.

\begin{figure}[!hbtp]
\includegraphics[width=8cm]{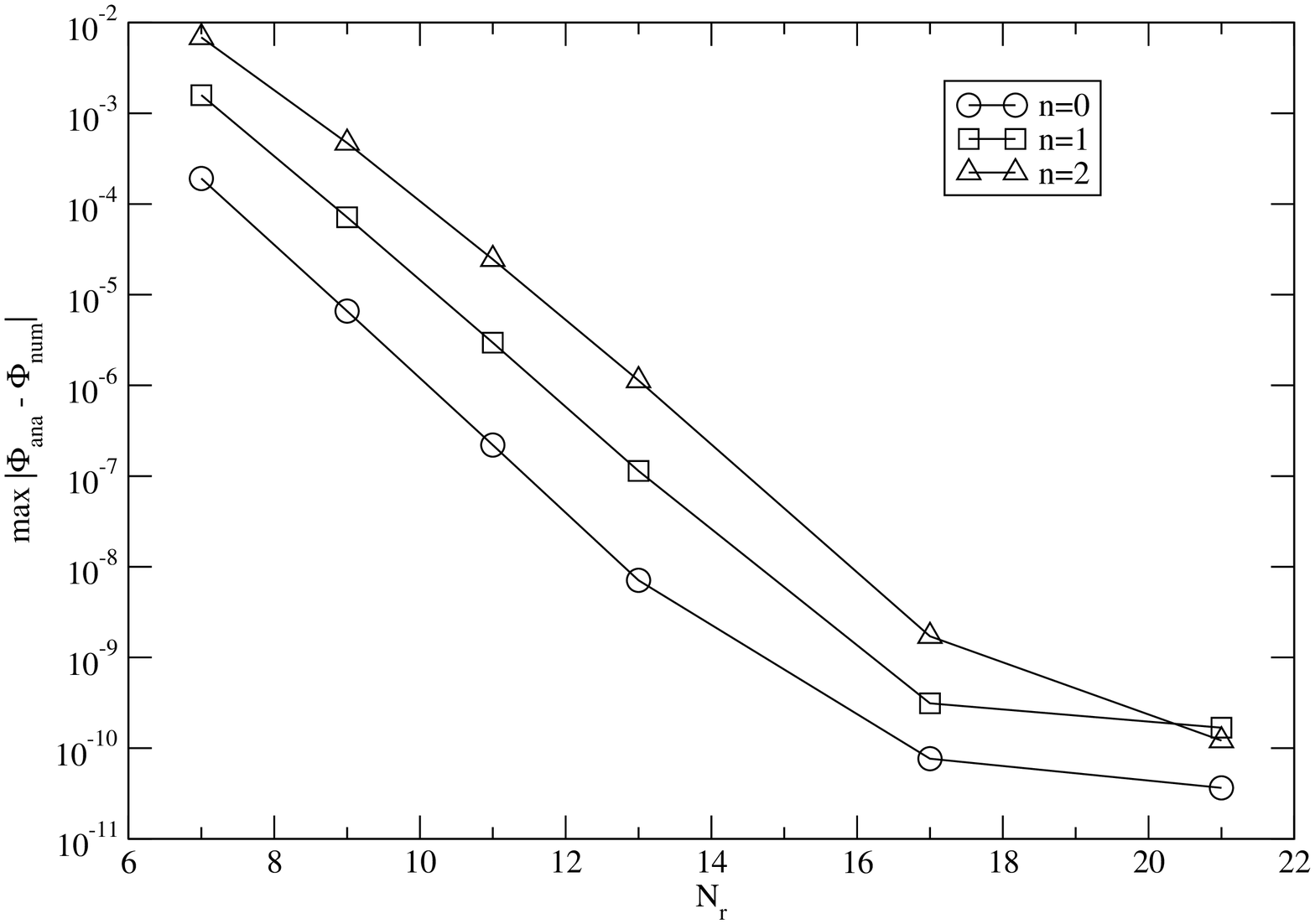}
\includegraphics[width=8cm]{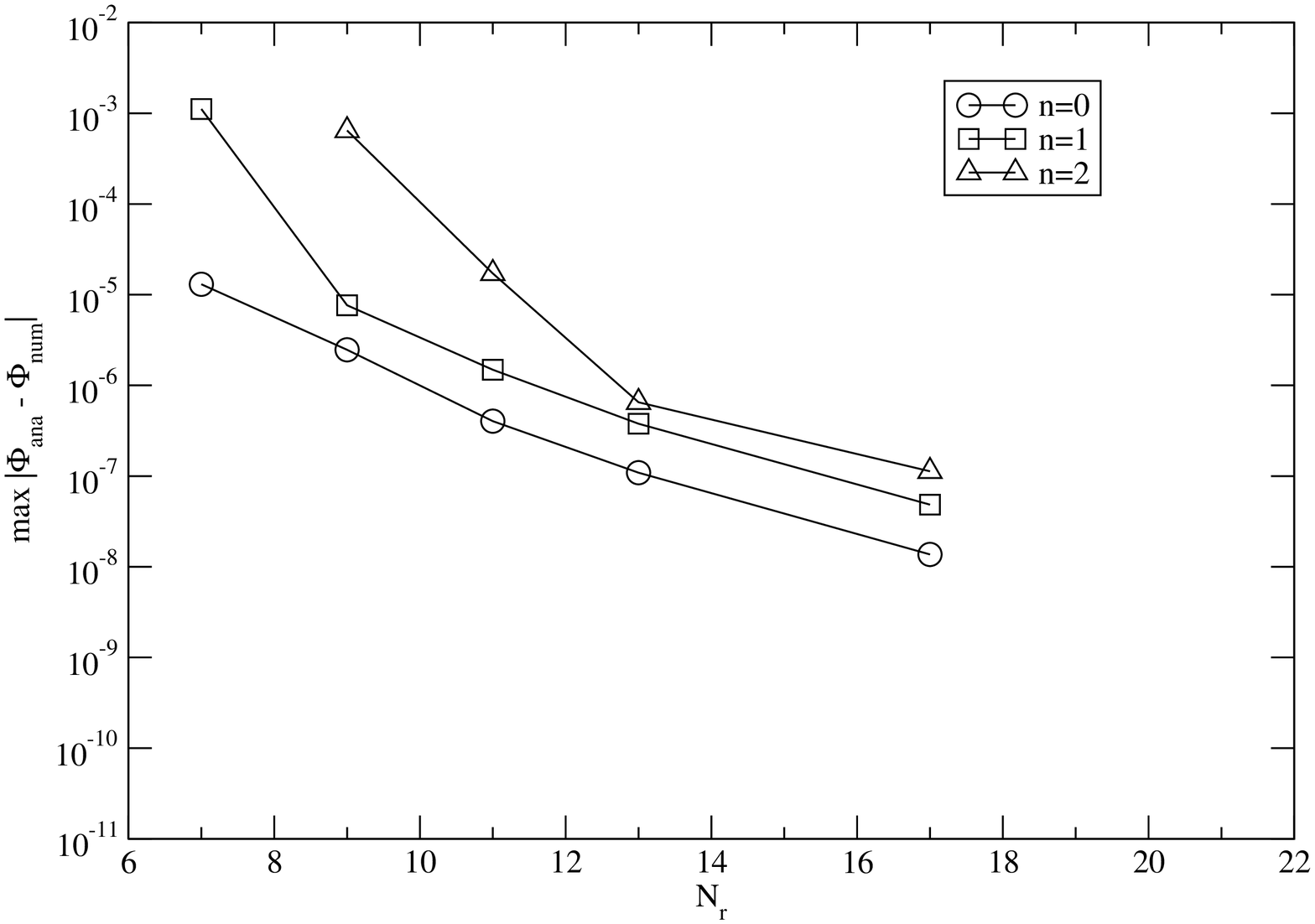}
\caption{Maximum difference between the analytical and numerical KG breathers (case $k=1$ and $m=1$). The first panel corresponds to Eq.~(\ref{fieldeqnum}) and the second one to  Eq.~(\ref{cfieldeq2num}). The number of temporal coefficients is fixed to $13$, and the results are shown as a function of the number of radial coefficients, for the solutions corresponding to $n=0$, $1$ and $2$. \label{figerrorlin}}
\end{figure}

Fig.~\ref{figerrorlin} shows the maximum difference between the numerical solution and the analytic one given by Eq.~(\ref{kgbreather}), in the case $k=1$. The first panel shows the results for the Schwarzschild area coordinates (Eq.~(\ref{fieldeqnum})) and the second one for the compactified ones (Eq.~(\ref{cfieldeq2num})). The results are obtained with a fixed number of points in the temporal direction ($N_t = 13$ to be precise) and one shows the result as a function of the radial number of points $N_r$ for the configurations with 0, 1 and 2 nodes.

Even if the absolute error is different in both cases, it decreases exponentially or faster, as expected with spectral methods, proving that the numerical solution of the equation is correct. On the left panel one can see a saturation at around $10^{-10}$ because it is the value of the threshold at which the iteration is stopped. It is true that in the linear case the solutions can be represented with only one coefficient in time (see Eq.~(\ref{oneharm})). However the number of coefficients in time is still important because it changes the convergence of the numerical algorithm to the solution. In particular, in order to increase $N_r$ one needs to also increase the temporal number of points. This effect explains why the curves of Fig.~\ref{figerrorlin} do not extend further in terms of $N_r$ ; it would require to also increase $N_t$.

\section{Self-interacting fields}\label{selfinter}

\subsection{The potential}

We assume that the scalar potential $U(\phi)$ has a zero valued minimum at $\phi=0$, and represent $U(\phi)$ by the scalar field mass $m$ and the expansion coefficients $g_j$
\begin{equation}
U(\phi)=\frac{1}{2}m^2\phi^2+\sum_{j=2}^\infty \frac{g_j}{j+1}\phi^{j+1} \  .
\label{phiexp}
\end{equation}
In order to be able to treat nontrivial massless potentials, such as $U(\phi)=\phi^4$, we do not take out a common $m^2$ factor from the $g_j$ coefficients in \eqref{phiexp}.

A general fourth order potential that is symmetric around $\phi=0$ can be written as $U(\phi)=a(\phi^2-b^2)^2+c$. Shifting the first minimum to the origin, this becomes $U(\phi)=a\phi^2(\phi-2b)^2$. Then the derivative is $U'(\phi)=8ab^2\phi-12ab\phi^2+4a\phi^3$, showing that the scalar field mass is $m^2=8ab^2$. Since the potential term is the only nonlinear term in the field equation, rescaling the field $\phi$ we can make $b=1$. If we use the compact coordinate system \eqref{confcord} then we can rescale $k$ in order to make $m=1$ in the field equation \eqref{cfieldeq2}. Using the Schwarzschild coordinates \eqref{schcoord} we also have to rescale $t$ and $r$, which changes the physical size and the frequency $\omega$ of the solution. Consequently, when considering a symmetric fourth order potential, from now on we will take
\begin{equation}
U(\phi)=\frac{1}{8}\phi^2(\phi-2)^2 \ . \label{phi4pot}
\end{equation}
The scalar field mass is then $m=1$ and the only nonvanishing coefficients in the expansion \eqref{phiexp} of the potential are $g_2=-\frac{3}{2}$ and $g_3=\frac{1}{2}$.

\subsection{Rescaled field variable}

Considering the asymptotic behavior of the Klein-Gordon breather solutions, it appears useful to introduce a rescaled field variable $\psi$ by
\begin{equation}
\phi=z^{\lambda_+/2}\psi=(\cos x)^{\lambda_+}\,\psi
=\frac{\psi}{(1+k^2r^2)^{\lambda_+/2}} \ ,  \label{psidef}
\end{equation}
where $\lambda_+$ has been defined in \eqref{lambdapm}. The variable $\psi$ tends to a time dependent, generally nonzero finite value at infinity. Let us take out the mass term from the potential, and denote the sum of the other terms by $\tilde U(\phi)$,
\begin{equation}
U(\phi)=\frac{1}{2}m^2\phi^2+\tilde U(\phi) \ . \label{utildedef}
\end{equation}
Then the field equation in conformal coordinates takes the form
\begin{equation}
 -\psi_{,\tau\tau}+\psi_{,xx}+\frac{D-1}{\sin x\cos x}\,\psi_{,x}-2\lambda_+\tan x\,\psi_{,x}=
\lambda_+^2\,\psi+\frac{1}{k^2(\cos x)^{\lambda_++2}}\tilde U'((\cos x)^{\lambda_+}\psi) \  , \label{psivar}
\end{equation}
while in Schwarzschild coordinates we have
\begin{align}
&-\frac{1}{1+k^2r^2}\psi_{,tt}+(1+k^2r^2)\psi_{,rr}
+\left[\frac{D-1}{r}+(D+1-2\lambda_+)k^2r\right]\psi_{,r} \notag\\
&\qquad=\frac{\lambda_+^2k^2}{1+k^2r^2}\psi
+(1+k^2r^2)^{\lambda_+/2}\tilde U'\left(\frac{\psi}{(1+k^2r^2)^{\lambda_+/2}}\right) \  .
\end{align}
All Fourier modes of $\psi$ tend to a generally nonzero finite constant at infinity. Except for the massless case in $D=1$ spatial dimension, the derivative of $\psi$ with respect to $x$ is zero at $x=\pi/2$. However, for nonzero $m$ the function $\psi$ is in general cannot be expanded in Taylor series with integer powers
at $x=\pi/2$, since its expansion contains $y^{n_1\lambda_++n_2}$ terms, where $y=\pi/2-x$, and $n_1$ and $n_2$ are nonnegative integers.

We have seen in Sec.~\ref{kleingordon} that all finite energy Klein-Gordon fields can be expanded as sums of the breather solutions $\phi^{(n)}_{KG}$. However, this is generally not true for self-interacting scalar fields, since the nonlinear terms in the field equation induce $y^{\lambda_+}$ terms into the expansion of the scalar field around $x=\pi/2$. Even for zero mass fields, when $\lambda_+=D$, the expansion in terms of $\phi^{(n)}_{KG}$ may fail when $D$ is odd. For example, for $D=3$ the expansion of $\phi^{(n)}_{KG}$ for all $n$ starts as $c_0y^3(1+c_1y^2+c_2y^4+\ldots)$, but if $g_2\neq0$, i.e.~the potential is not symmetric around its minimum, then the quadratic term in the field equation induces $y^6$ terms in $\phi$, which are absent in the expansion of the Klein-Gordon breathers.

\subsection{Periodic solutions} \label{persolsec}

In this section we present numerical solutions to the nonlinear KG equation with the potential Eq. (\ref{phi4pot}). As in Sec. \ref{spectral} the Kadath library is used with the same setting. In addition to the two equations already used in linear case (Eqs. (\ref{fieldeqnum}) and (\ref{cfieldeq2num})), we also solved the system using the rescaled variable $\psi$. The numerical equation used is Eq. (\ref{psivar}) rewritten as
\begin{equation}
\cos^2 x \l(-\psi_{,\tau\tau}+\psi_{,xx}\r)+ \l(D-1\r) \frac{\cos x}{\sin x} \psi_{,x} - 2\lambda_+ \sin x \cos x \psi_{,x} = \lambda_+^2 \cos^2 x \psi
+ \frac{1}{k^2}\psi \l(-\frac{3}{2} \phi + \frac{1}{2} \phi^2\r),  \label{psivarnum}
\end{equation}
where $\tilde{U}'$ has been replaced by its explicit value given by Eq.~(\ref{phi4pot}). As usual, the term $\psi_{,x}/\sin x$ is computed in the coefficient space to avoid trouble near $x=0$. At infinity the equation is degenerate but one can easily see that the appropriate boundary condition (given by regularity arguments) is $\psi_{,x}\l(x=\pi/2\r)=0$. This illustrates the fact already mentioned that $\psi$ does not vanish at infinity.

Contrary to the linear case, solutions no longer exist only for discrete sets of values of $\omega$ but rather for whole intervals of frequency. The linear solutions can be used as good initial guess to the non-linear case when the amplitude is small. The way to proceed is the following. Set the frequency to a value close to the linear case : $\omega = \omega_{\rm lin} + \delta_\omega$ and scale the linear solution so that its value at the origin is small and given by $\delta_\phi$. For each choice of $\l(\delta_\omega, \delta_\phi\r)$, one can launch the Newton-Raphson iteration of Kadath. If the values are not good enough (for instance if $\delta_\omega$ has the wrong sign or is too big), then the code either fails to converge or reaches the trivial solution $\phi=0$ everywhere. Nevertheless, it is usually easy to find values so that the code finds a true solution of the non-linear problem. Once a solution is found, one can then slowly vary the frequency to construct whole sequences of solutions.

Most of the data presented in this section are obtained using the formalism of Eq.~(\ref{psivarnum}). Checks have been made to ensure that the results are the same when using the other formulations (\ref{fieldeqnum}) and (\ref{cfieldeq2num}). The standard numerical setting makes use of 4 domains and one uses $N_r=33$ radial coefficients and $N_t=17$ temporal ones. The Newton-Raphson iteration is stopped when the error on the equations reaches a threshold of $10^{-8}$. In the following, the influence of both the formulation and the resolution is shown in a few chosen cases to ensure that the observed behaviors are real.

\begin{figure}[!hbtp]
\includegraphics[width=8cm]{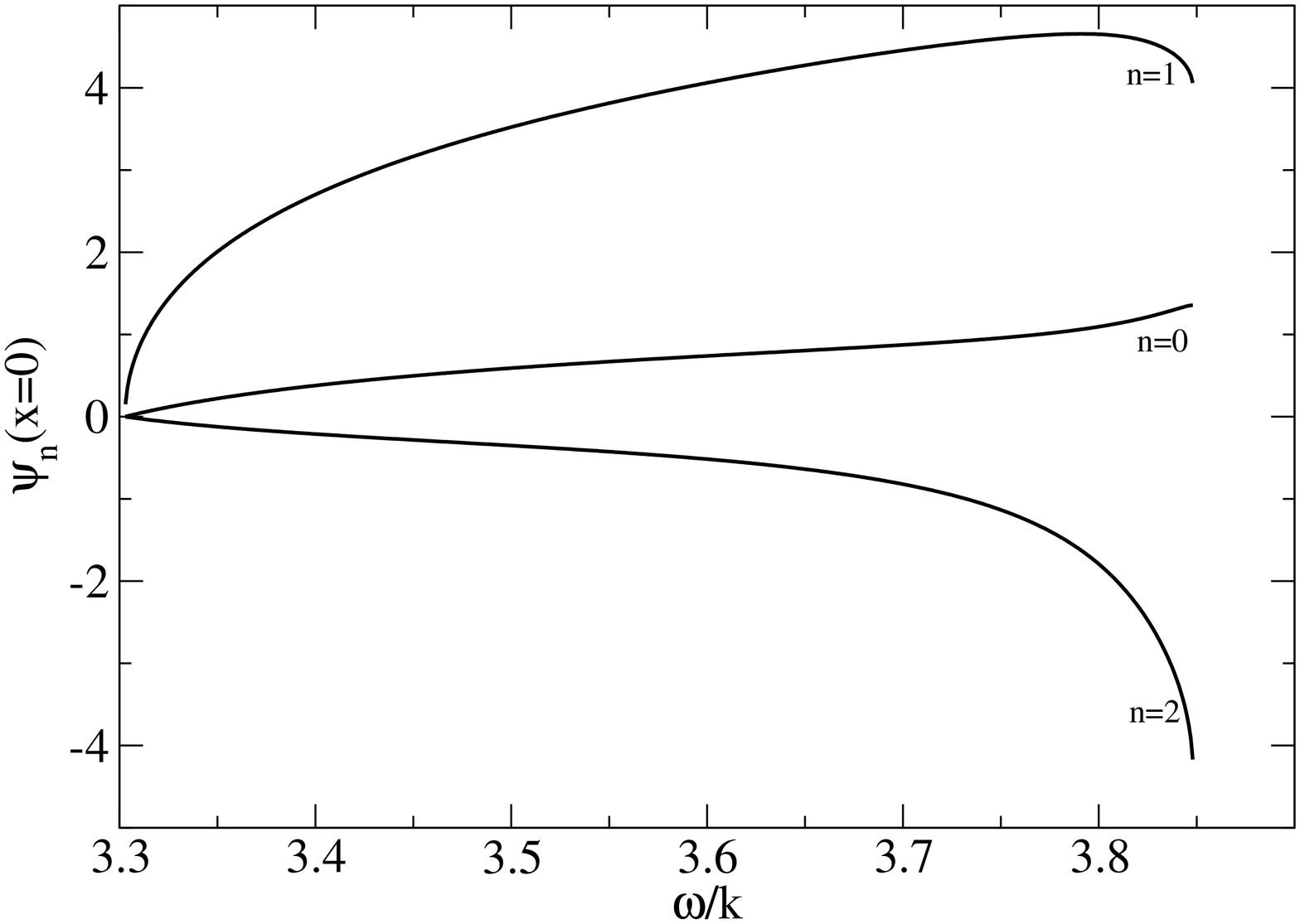}
\includegraphics[width=8cm]{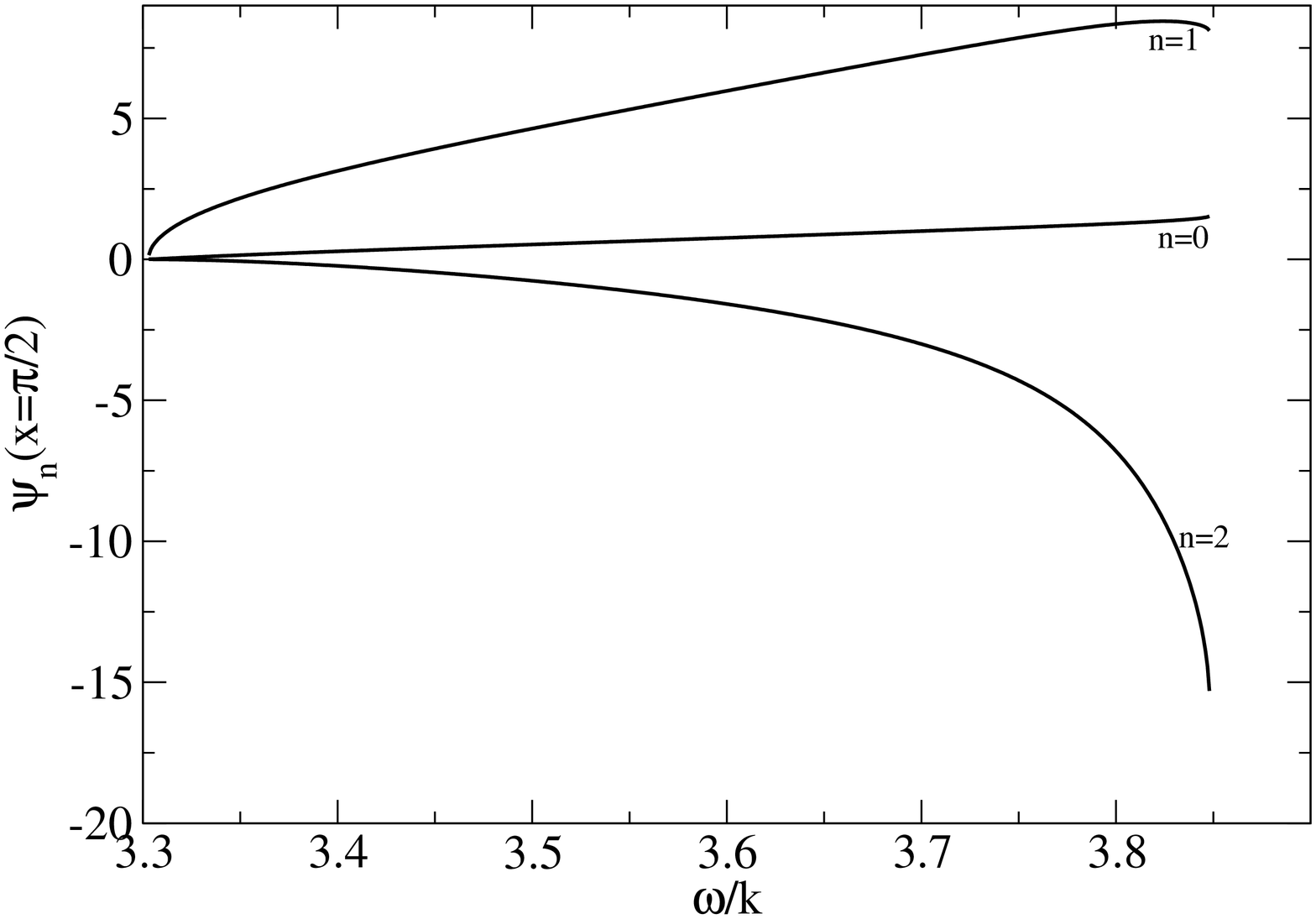}
\caption{ For $k^2=1$, the first panel shows the value of the first modes at the origin (i.e. $\psi_n \l(x=0\r)$). The second panel shows the value of the modes at infinity (i.e. $\psi_n \l(x=\pi/2\r)$).\label{valsk1}}
\end{figure}

This paper does not present an exhaustive exploration of the parameter space. We present results for $m=1$, $D=3$ and for three different cosmological constants $k^2=1$, $k^2=0.1$ and $k^2=0.001$. Moreover, we concentrate on solutions that tend to the nodeless fundamental Klein-Gordon breather in the linear regime. Even in those few selected cases, the structure of the parameter space is already rich and complex. For $k^2=1$ the values of the first modes $\psi_n$ are shown on Fig. \ref{valsk1}, both at the origin $x=0$ (first panel) and at infinity $x=\pi/2$ (second panel). Let us mention that, since the potential (\ref{phi4pot}) is not symmetric around its minimum, the signs of all odd modes can be reversed to get another perfectly valid solution, which is just a half-period phase shifted version of the original. One can also note that the dominating mode is $n=1$ which is the only one non-zero in the linear case. Finally, as expected, one can see that all the modes vanish when $\omega/k$ tends to its value for the linear KG breather.

\begin{figure}[!hbtp]
\includegraphics[width=8cm]{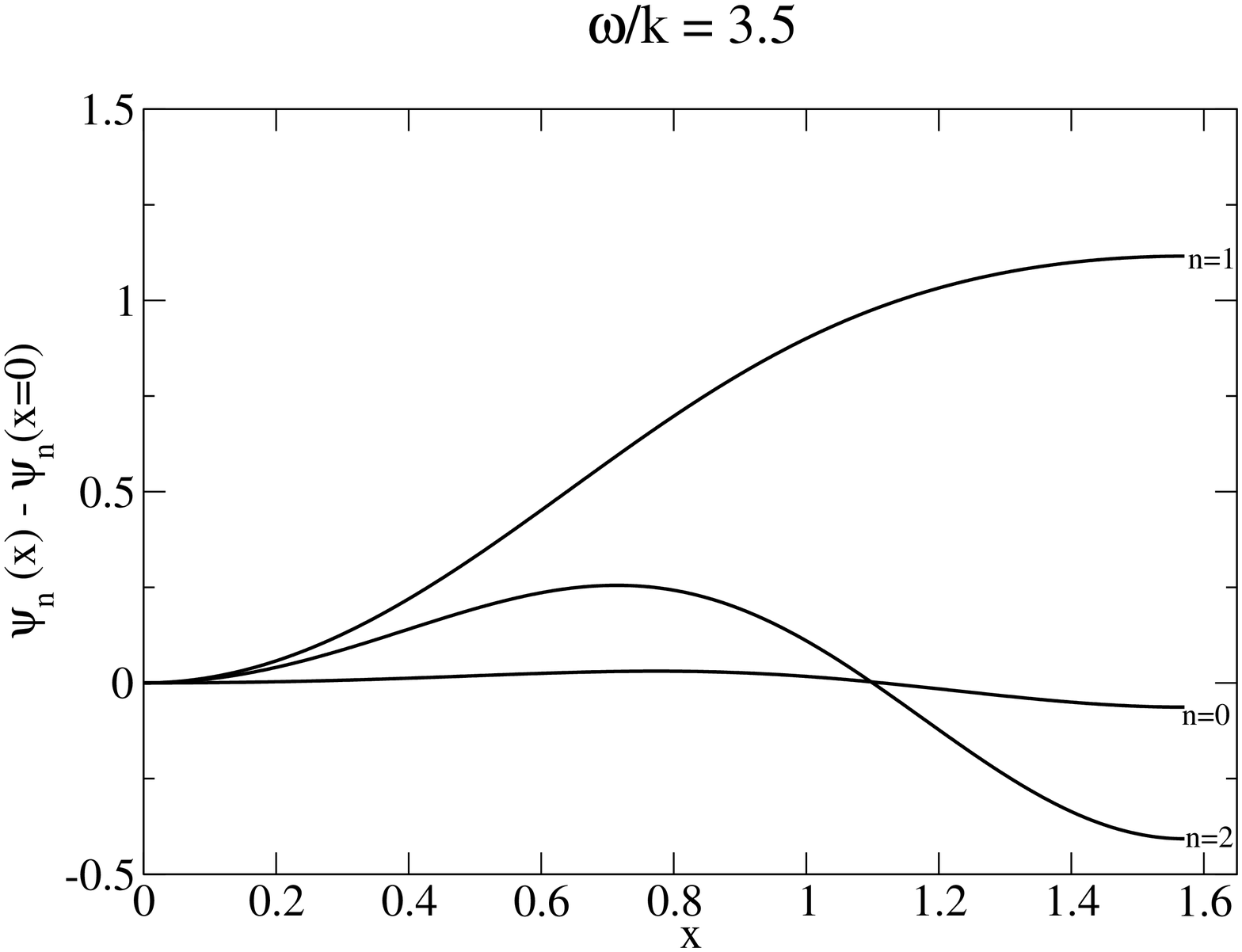}
\includegraphics[width=8cm]{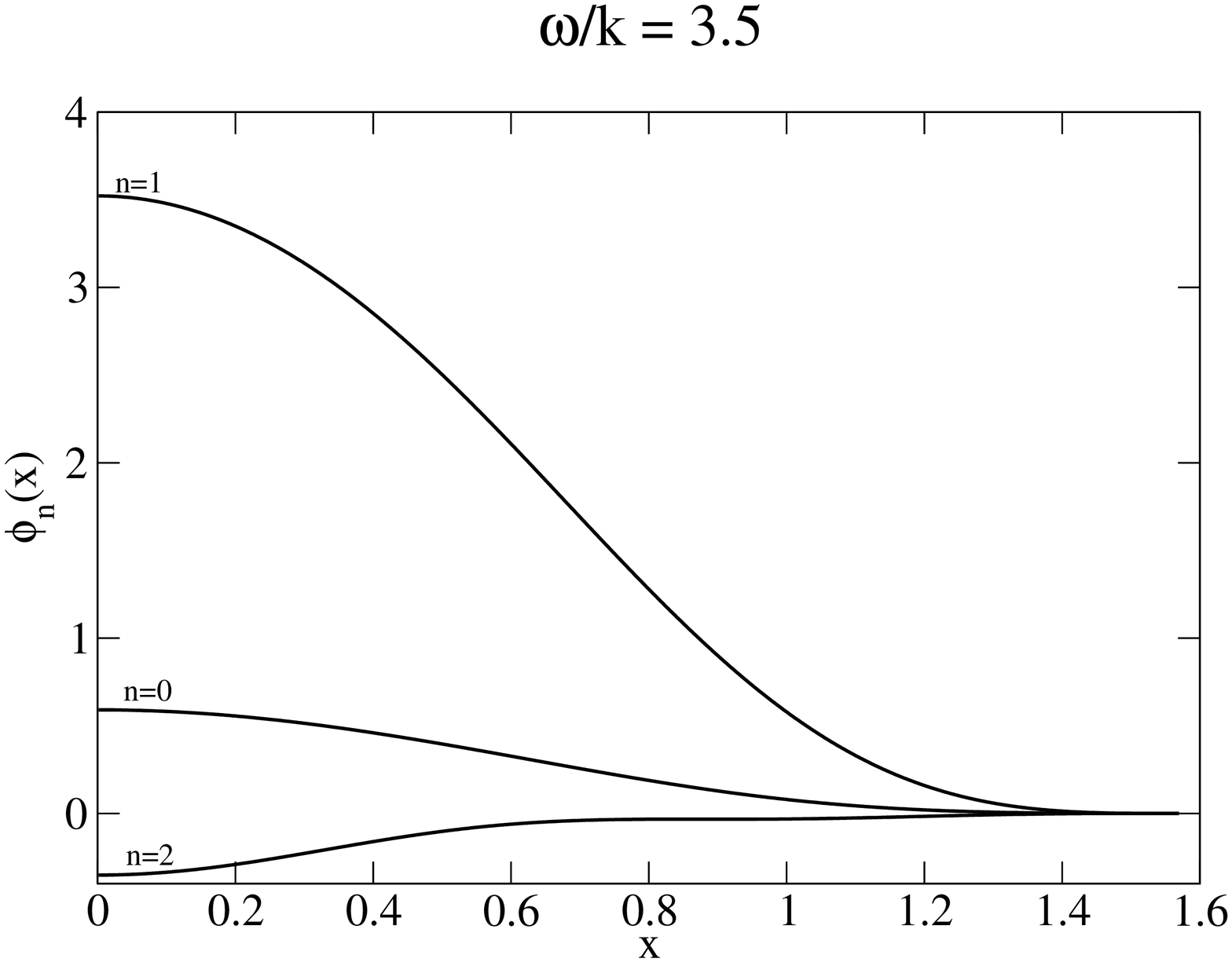}
\caption{ Profiles of the first three modes for $k^2=1$ and $\omega/k=3.5$. The first panel shows the modes in terms of the variable $\psi$ and the second one in terms of the ``true'' field $\phi$. \label{profs3.5}}
\end{figure}

Figure \ref{profs3.5} shows the radial dependence of the first three modes for $k^2=1$ and $\omega/k=3.5$. The first panel shows the result in terms of the auxiliary variable $\psi_n$ (shifted so that the curves vanish at the origin), while the second panel shows the value of the ``true'' field $\phi_n$. As expected, the variable $\psi_n$ does not go to zero at infinity.

\begin{figure}[!hbtp]
\includegraphics[width=8cm]{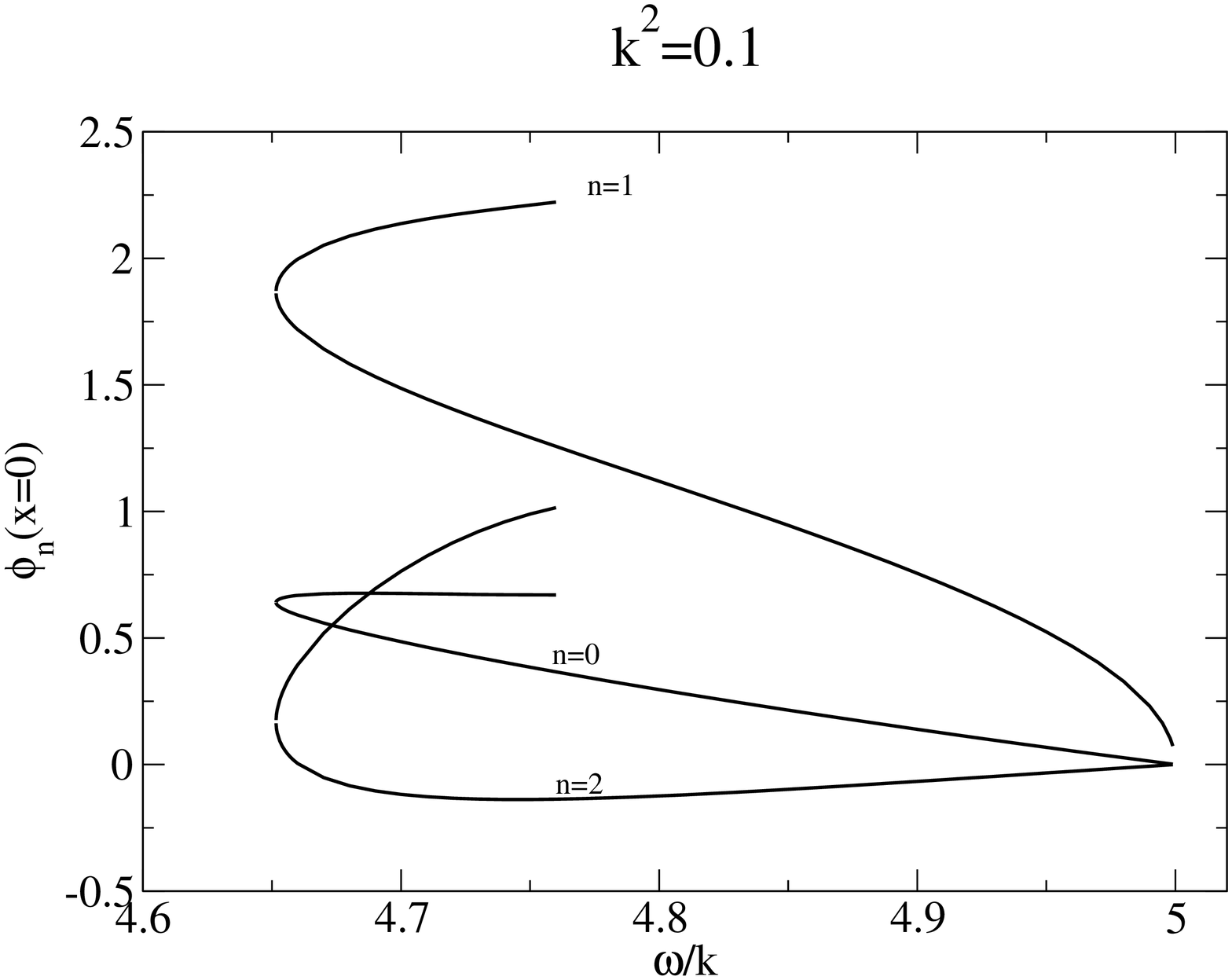}
\includegraphics[width=8cm]{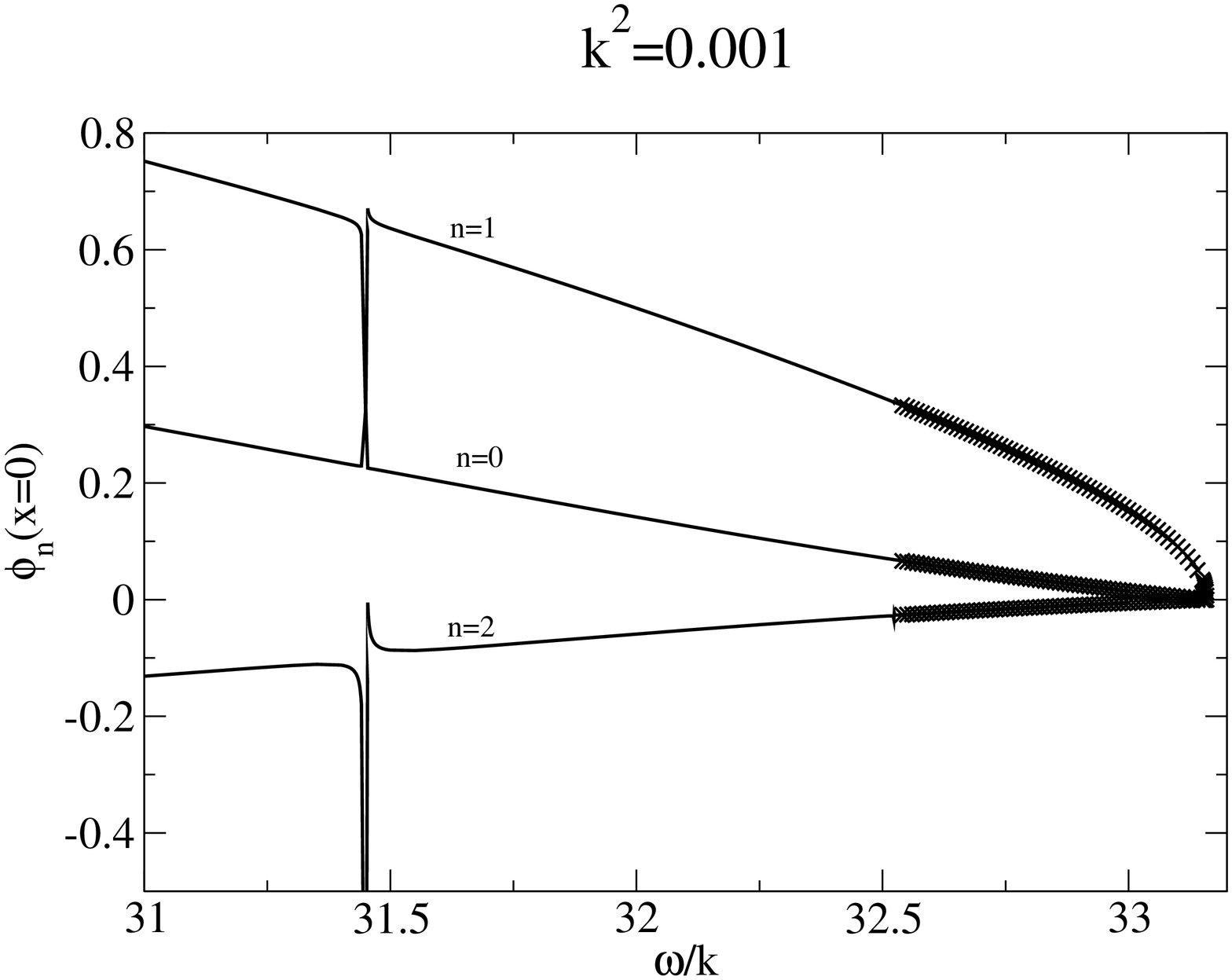}
\caption{Values of the first three modes at the origin, as a function of $\omega/k$ for $k^2=0.1$ (first panel) and $k^2=0.001$ (second panel). In this last case, the solid lines denote the results obtained using the variable $\phi$ and the crosses the variable $\psi$. \label{orik}}
\end{figure}

As an illustration of the fact that the various solutions can be very different when changing $k^2$, one shows the value of the first modes at the origin as a function of $\omega$ for $k^2=0.1$ and $k^2=0.001$ on Fig.~\ref{orik}. In the case $k^2=0.1$, we observe a turning point for $\omega/k \approx 4.652$ and no solutions are found below this frequency. It will be seen in Sec.~\ref{timeevol}, that in this case there is a dynamical instability that sets in before this turning point so that the configurations corresponding to the upper branch (as some on the lower branch) are unstable.

For the case $k^2=0.001$ no such turning point is observed but there are nevertheless some interesting effects. First, for this value of the cosmological constant, it turns out that the numerical code using the variable $\psi$ is less stable than the one using the standard field $\phi$. As will be seen later, a possible explanation is the behavior of the modes $\psi_n$ at infinity. Let us note that in the cases where both codes converge they do lead to the same results. This is clearly seen in the second panel of Fig. \ref{orik}, where the results obtained by the variable $\psi$ (crosses) lie exactly on top of the ones obtained via the variable $\phi$ (solid lines).

\begin{figure}[!hbtp]
\includegraphics[width=8cm]{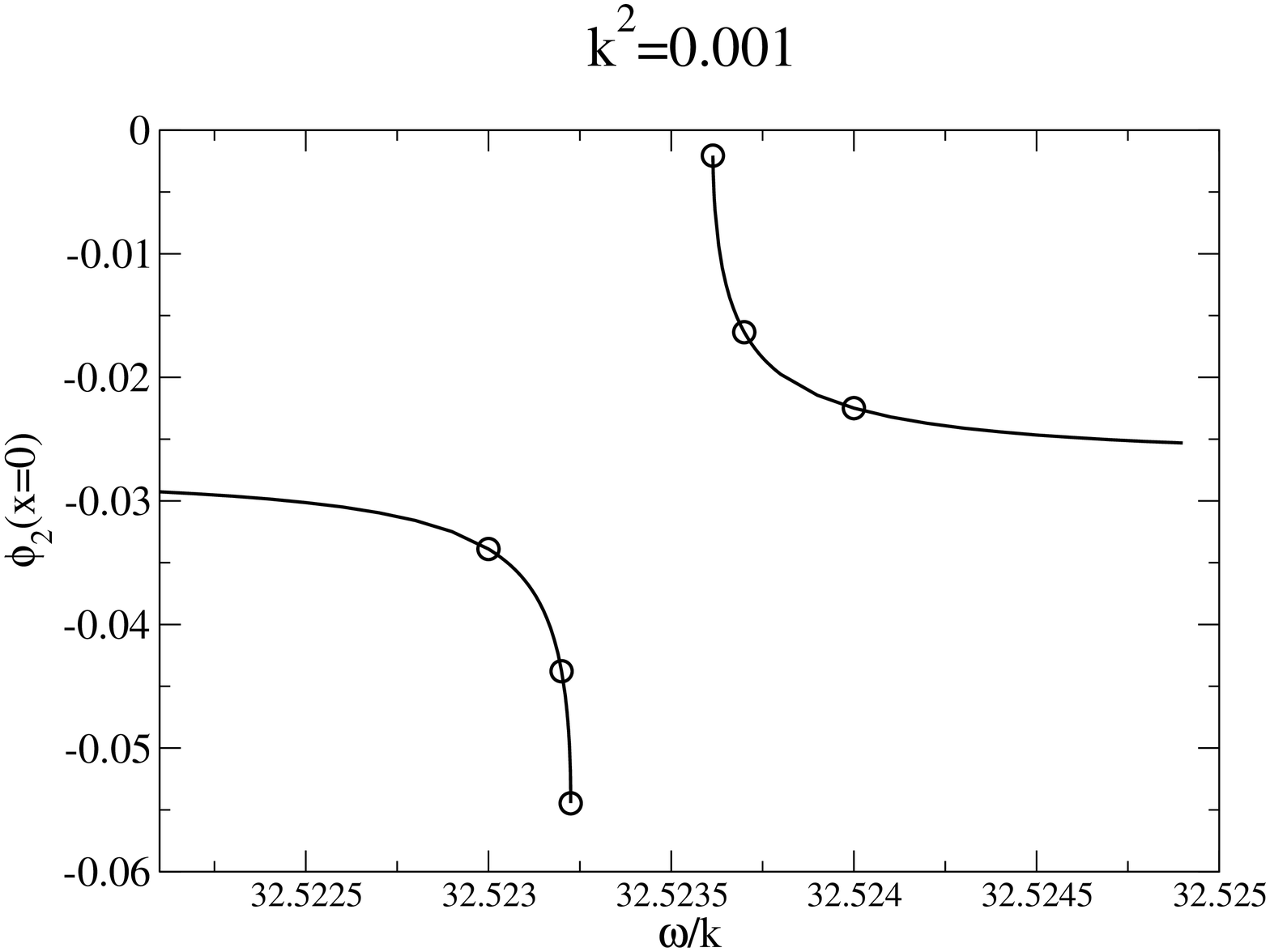}
\includegraphics[width=8cm]{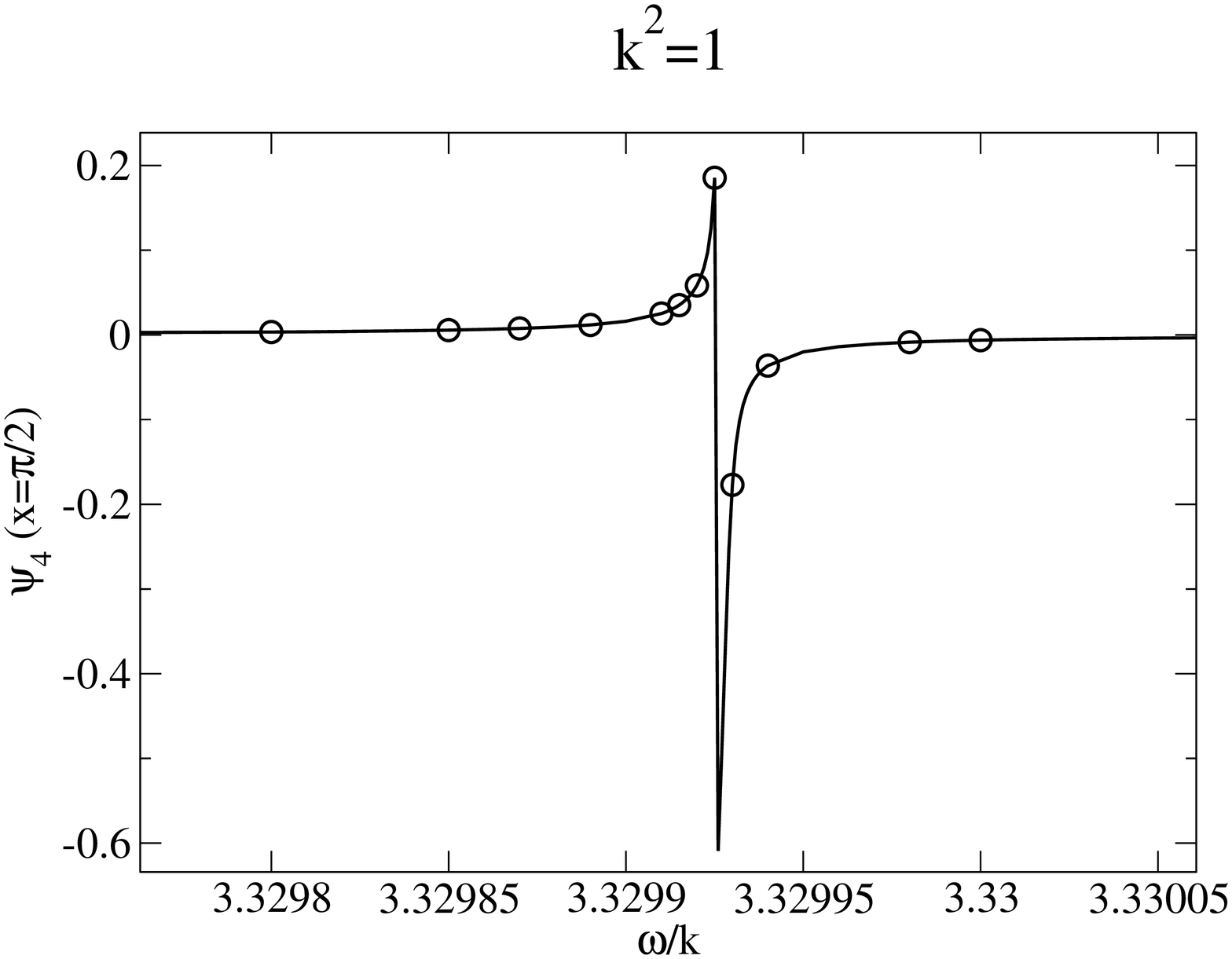}
\caption{The first panel shows the value of $\phi_2$ at the origin for a resonance with $k^2=0.001$. The second panel shows the value of $\psi_4$ at $x=\pi/2$ as a function of $\omega/k$ around a resonance for $k^2=1$. In both cases, the solid line denotes the results obtained with the standard resolution and the circles the ones with the higher resolution. \label{reson_std}}
\end{figure}

A striking feature of the case $k^2=0.001$ is the presence of resonance-like behavior, where the values of the various modes change very rapidly. The most evident case is for frequencies just below $\omega/k=31.5$ but this is not the only one. A closer examination of the data reveals the presence of another resonance around $\omega/k=32.5$, close to the frequency where the code using $\psi$ starts to fail. This is shown on the first panel of Fig. \ref{reson_std}, where the value of $\phi_2$ at the origin is plotted as a function of $\omega/k$. We show this mode because it seems to be the one for which the resonance is the strongest. In order to verify the reality of the observed behavior, we checked that the results are not changing with resolution. The circles denote configurations obtained with $N_r=49$
 and $N_t=21$ and they fall nicely on the curve obtained with the standard setting. It turns out that the resonance-like behavior is not confined to small values of $k^2$ and is also found in the $k^2=1$ case. This is illustrated by the second panel of Fig. \ref{reson_std} where the value of $\psi_4$ at $x=\pi/2$ is shown. There are however some differences: in the case $k^2=0.001$ the effect of the resonance is stronger for the mode $\phi_2$ whereas it is stronger for $\psi_4$ in the case $k^2=1$. Also in the case $k^2=0.001$, Fig. \ref{reson_std} seems to indicate a gap in frequency between the two sides of the resonance. This effect is not seen for $k^2=1$.

\begin{figure}[!hbtp]
\includegraphics[width=8cm]{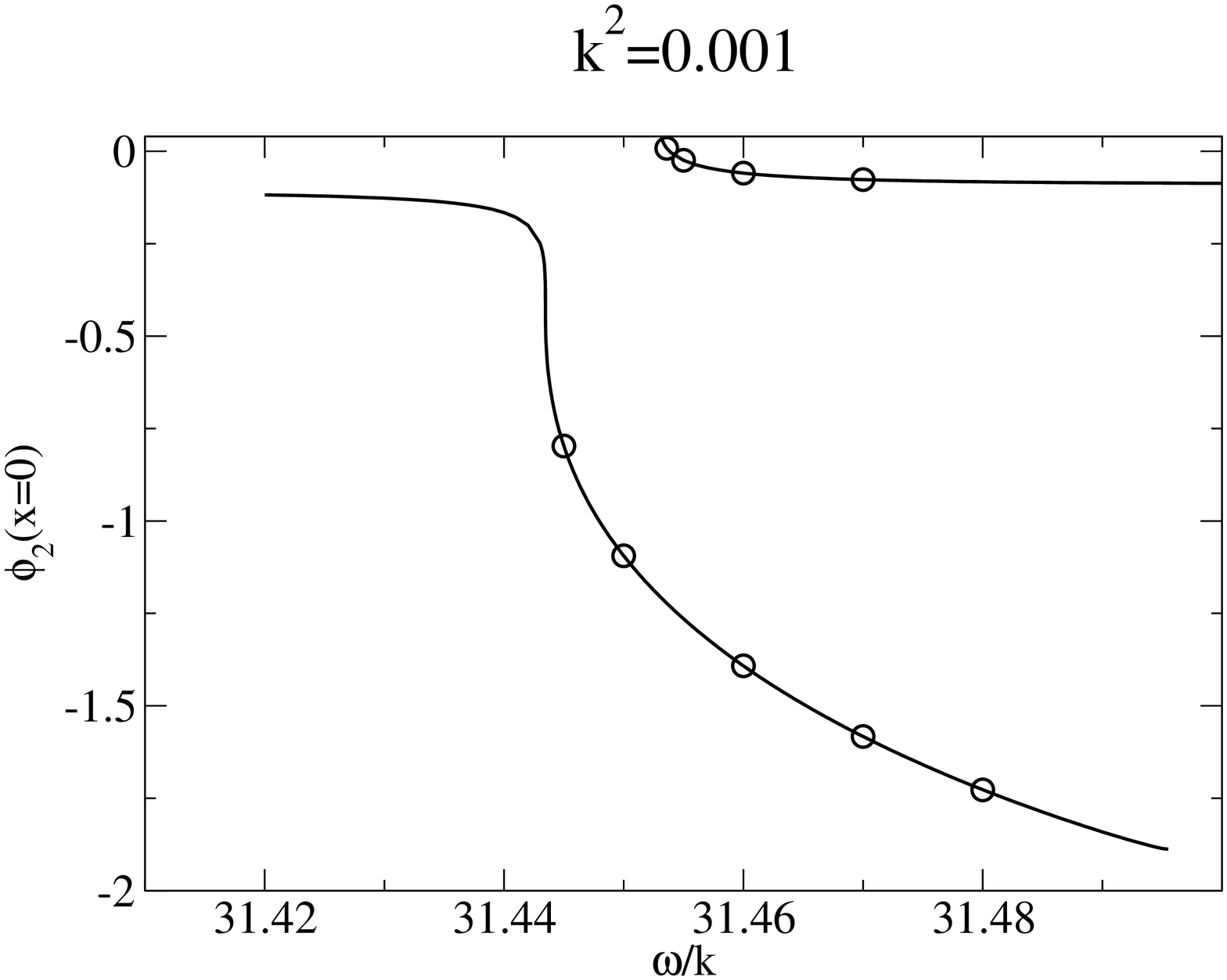}
\includegraphics[width=8cm]{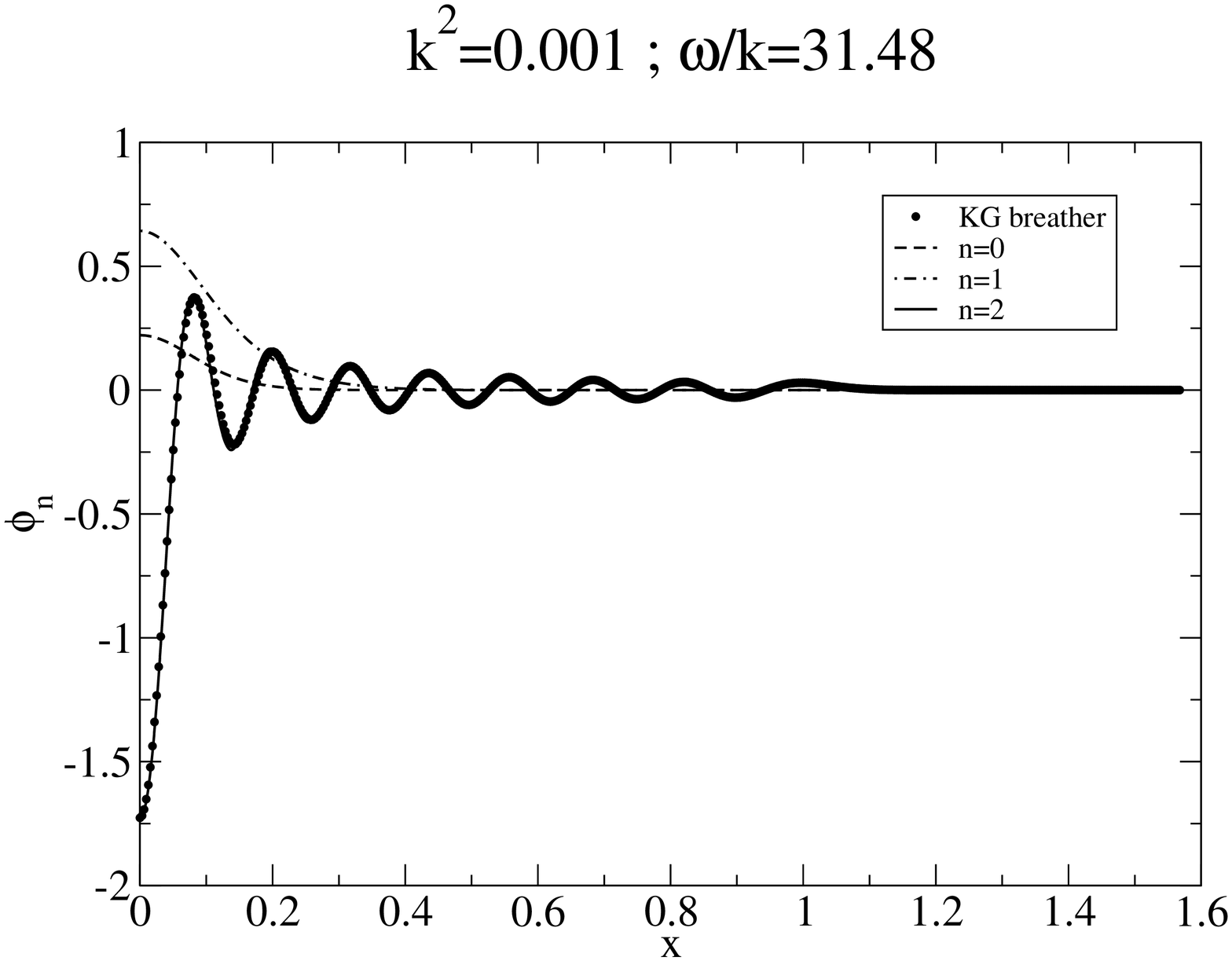}
\caption{The first panel shows the value of $\phi_2$ at the origin around a resonance for $k^2=0.001$. The circles denote the results obtained with the higher resolution. The second panel shows the structure of the various modes for a configuration of the lower branch ($\omega / k = 31.48$). The mode $n=2$ is very close the breather with 15 nodes, which is shown by the solid circles. \label{reson_2}}
\end{figure}

Even at fixed cosmological constant, the structure of the resonances can vary from one another. In the first panel of Fig. \ref{reson_2}, the value of $\phi_2$ at the origin is shown, for the second resonance observed at $k^2=0.001$. In this case, there seems to be two branches of solutions and we do find values of $\omega/k$ for which two configurations coexist. The numerical code converges to one branch or the other, depending on the initial guess (i.e. depending if one proceeds by increasing or decreasing $\omega/k$).

A closer examination of the structure of the various modes lead us to formulate a plausible explanation for the resonance-like behaviors. As seen in the second panel of Fig. \ref{reson_2}, one of the modes, here $n=2$, appears to decouple from the other ones. Its shape is indeed very close to the breather with 15 nodes, which is also shown on the plot. The frequency of this breather is $\omega / k \approx 63.158$, which is close to the double frequency of the solution $2\times 31.48 = 62.96$. A similar explanation seems to hold for the other resonances we observe: one of the modes starts to decouple from the others and gets very close to one of the Klein-Gordon breather solutions. This appears when the frequencies of the breather and of the non-linear solution are almost in rational ratios, as typical for a resonance-like behavior. If this explanation holds, other resonances could also exist that we did not observe. It is unclear at this point if this comes from the fact that the resonances are not present due to some non-linear effects or if they are so narrow that it is difficult to find them numerically. However an extended study of those effects is beyond the scope of this paper.

\begin{figure}[!hbtp]
\includegraphics[width=8cm]{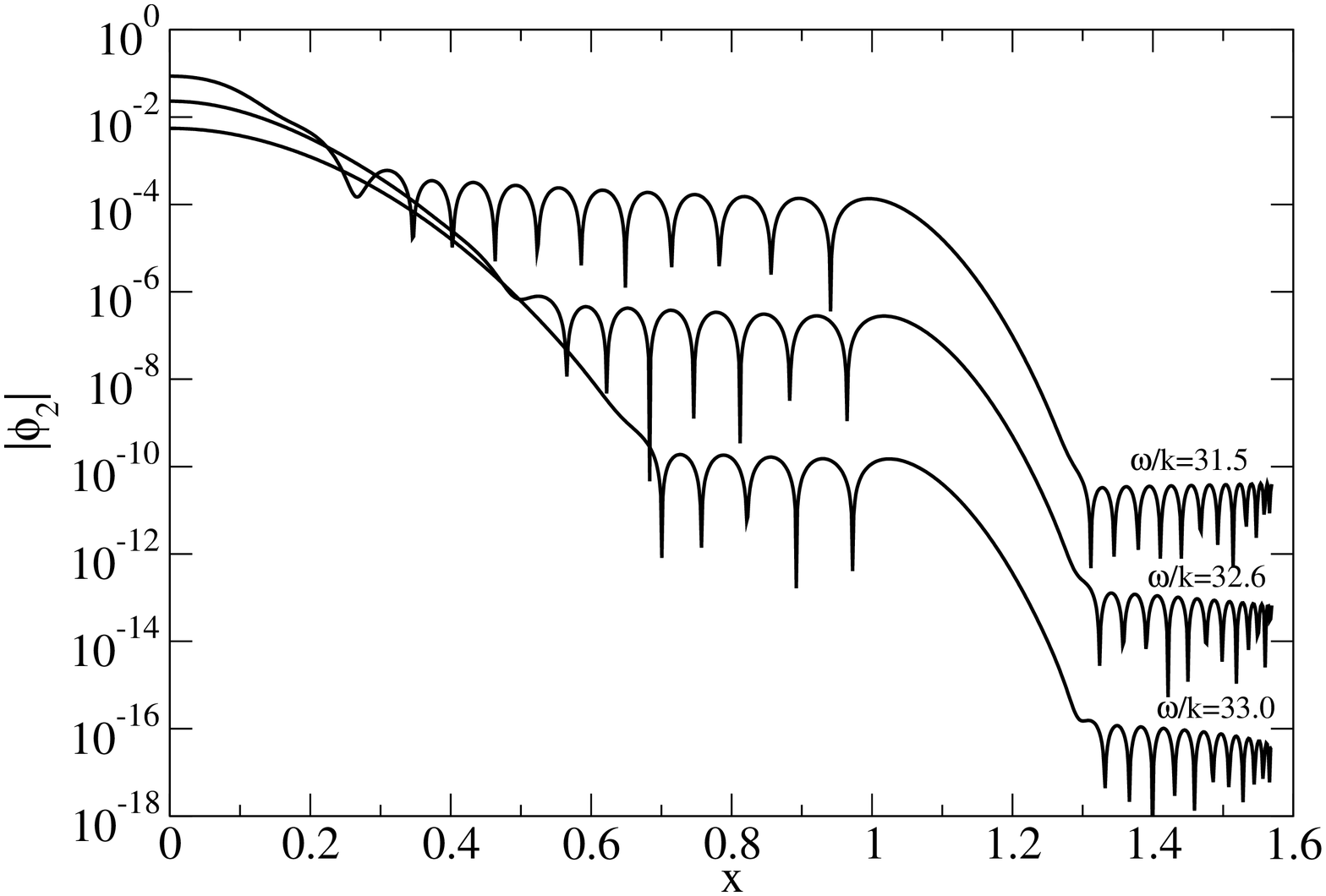}
\includegraphics[width=8cm]{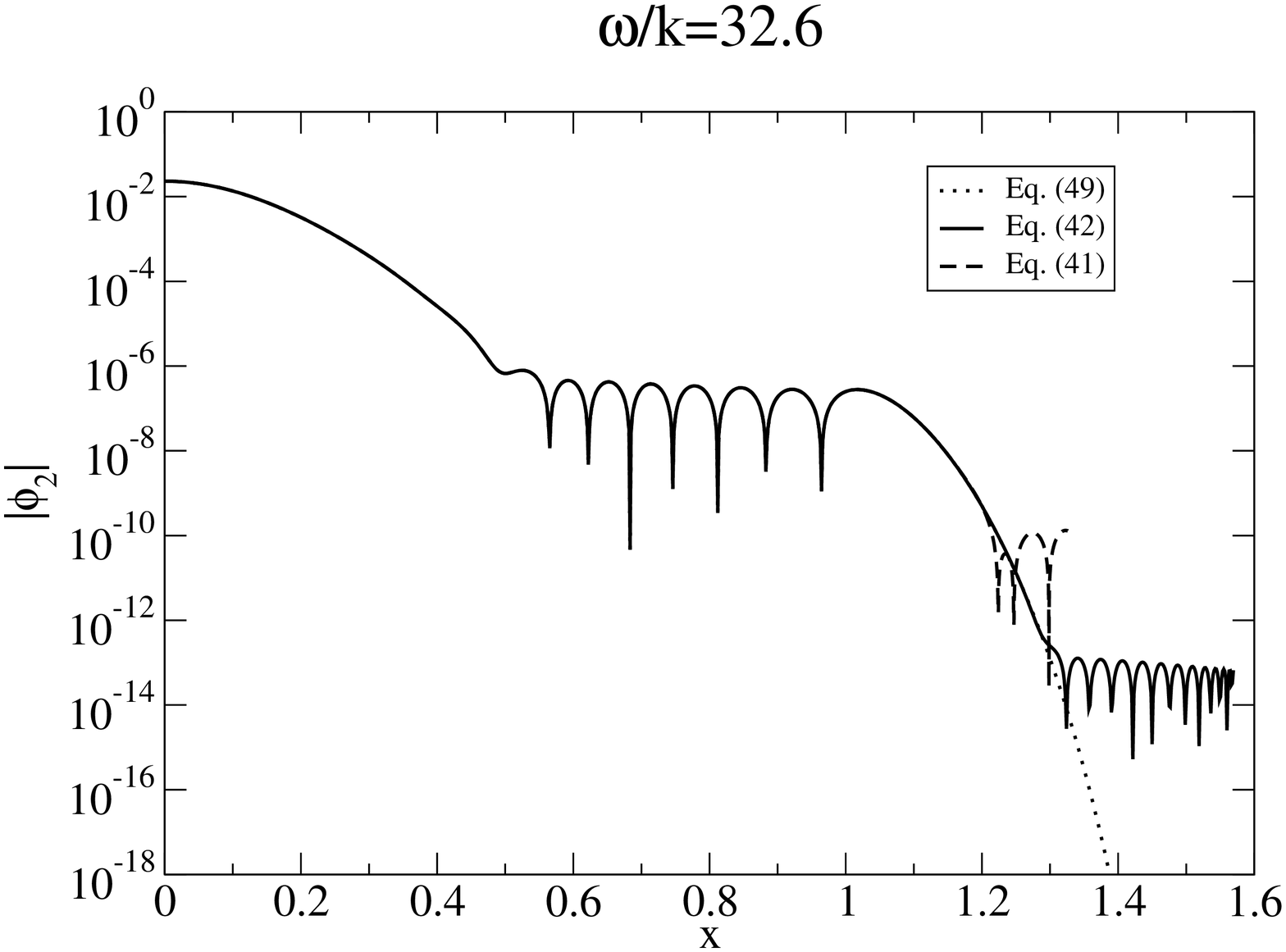}
\caption{Structure of the mode $\phi_2$ for $k^2=0.001$ for three different values of $\omega/k$ (first panel), and for three different numerical codes and $\omega/k=32.6$ (second panel). \label{profoscill}}
\end{figure}

Another interesting feature of the solutions for $k^2=0.001$ is to be observed in the radial dependence of the fields themselves. For such low values of the cosmological constant, the modes, starting from $n=2$, start to exhibit an oscillating structure. This is shown on the first panel of Fig. \ref{profoscill}, for three different values of $\omega$. In order to check the reality of those oscillations, it has been verified that, apart from the distant region ($x>1.2$), the field's profile  did not vary when changing the numerical setting (boundaries of the various domains, number of coefficients, or system of equations used). An illustration of that can be seen in the second panel of Fig.~\ref{profoscill} where the result for $\phi_2$ is shown for the three different numerical codes based on Eqs.~(\ref{fieldeqnum}, \ref{cfieldeq2num}, \ref{psivarnum}). The fact that the oscillations in the range $x \in \l[0.7, 1.1\r]$ are the same in all cases indicates that they are real and not numerical artefacts. This is to be contrasted with what is observed at very high radius (i.e. for $x>1.2$) where the solution is dominated by numerical noise (note the very small amplitude of the field at those large distances).

Schematically the mode $\phi_2$ (and also the higher order ones) can be decomposed into three regions : i) a core for $x<0.5$, ii) an oscillating regime for $x \in \l[0.5, 1.\r]$ and iii) a rapidly decaying part for $x>1$. The boundaries between the various regimes are given as an illustration for they depend on the frequency. The reason for the appearance of the oscillating regime can be found by recalling that one is looking at solutions for very low cosmological constant. This means that its effect is noticeable only at large distances. In the inner regions, it is like if the field was living in flat spacetime. In a previous paper, we studied precisely the structure of a self-interacting field in flat spacetime \cite{oscillons} and showed that indeed the mode $n=2$ and the subsequent ones had to develop an oscillatory tail, very similar to the one seen in Fig. \ref{profoscill}. The only difference is then observed at large distances where the effect of the cosmological constant starts to be noticeable and causes a damping of the field.

\subsection{Time evolution}\label{timeevol}

In order to check our methods we use the results of the spectral code as initial data for a spherically symmetric time evolution code. Since the initial data necessarily contains some numerical error, this also provides a good way to check the stability of the given solution. The time evolution code is a slightly modified version of the fourth order method of line code developed in \cite{FR,FR2}, and used for studying oscillons in \cite{oscillons,FFHL2,FFHM,moredim,dilaton}. All configurations obtained by the spectral code for $k=1$, which can be seen on Fig.~\ref{valsk1}, evolve periodically to a very high precision, which indicate that they are stable. The amplitude remains constant to six degrees for several thousands of oscillations, and the frequency agrees with the frequency chosen in the spectral code to similar precision.

For $k^2=0.1$ there are already unstable solutions among the states represented on the first panel of Fig.~\ref{orik}. Small amplitude solutions are stable, but as the central amplitude grows the frequency decreases, and  we get to a point at $\omega/k=4.6689$ where instability sets in. Below this frequency all states are unstable in lower branch on the first panel of Fig.~\ref{orik}, down to $\omega/k=4.6515$, where the frequency becomes increasing with increasing amplitude. All states that we could construct with the spectral code on the upper branch are also unstable. In the whole studied domain the central energy density and the total energy are monotonically increasing functions of the central amplitude, and hence the onset of instability does not coincide with an energy maximum, as generally happens in the asymptotically flat case.

\begin{figure}[!htbp]
\includegraphics[width=10cm]{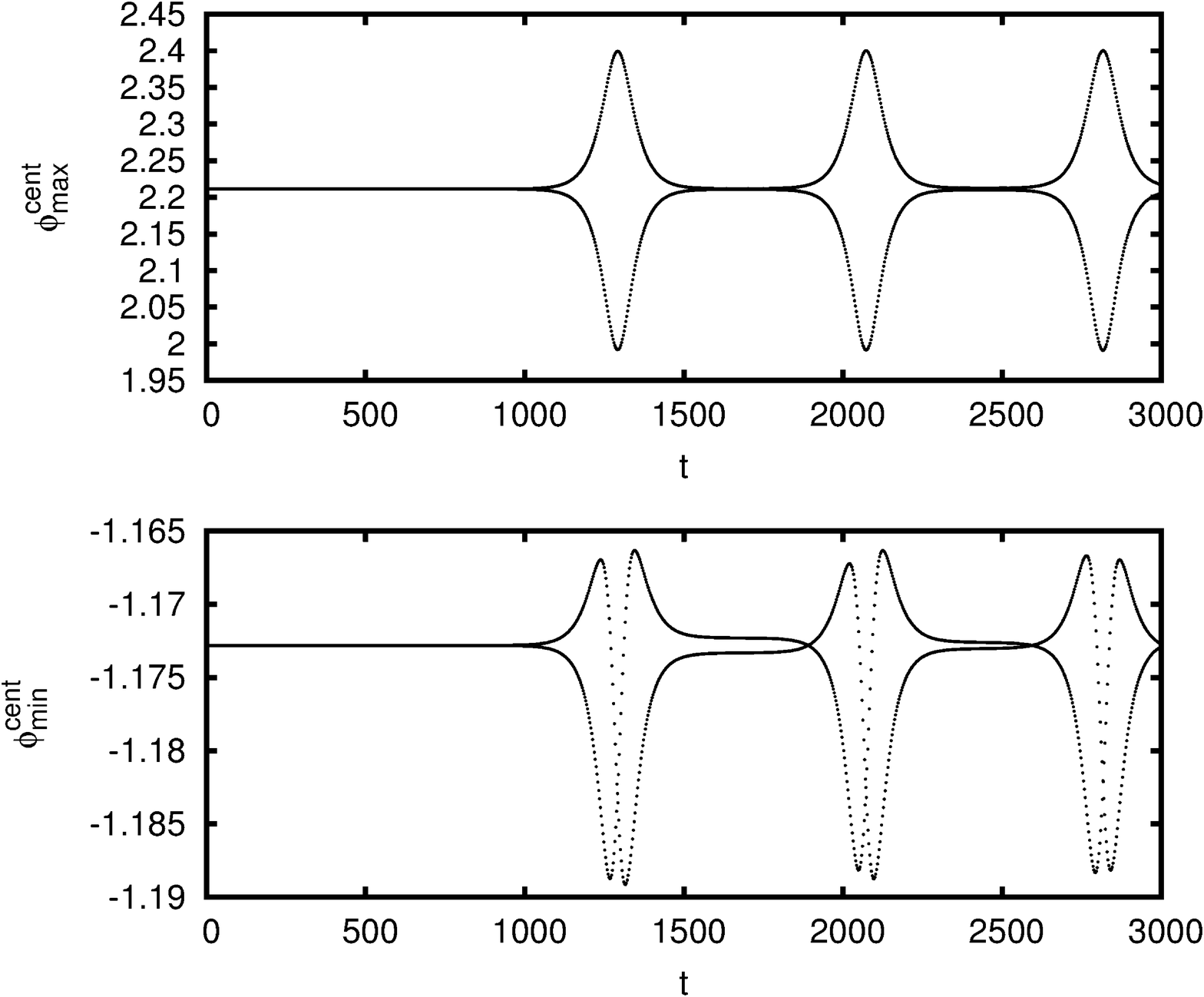}
\caption{Maximum and minimum values of the oscillating scalar field $\phi$ at center for an initial data with $\omega/k=4.6680$. The odd and even points follow different curves. \label{minmaxfig}}
\end{figure}

\begin{figure}[!htbp]
\includegraphics[width=10cm]{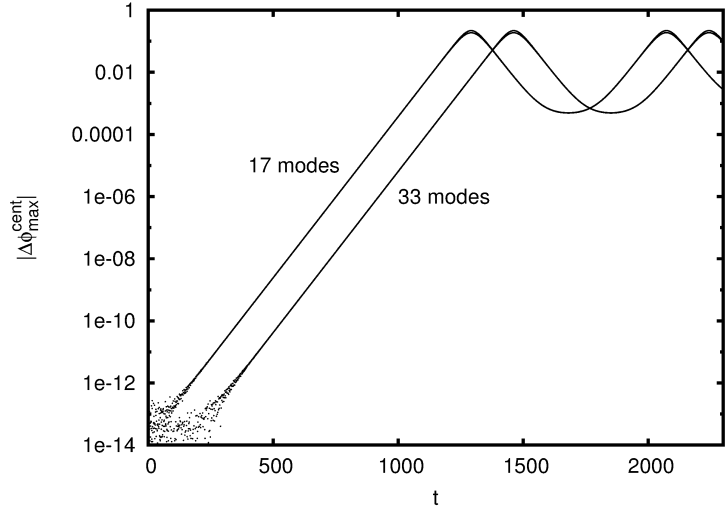}
\caption{Time evolution of the absolute value of the difference of the maximums of the scalar field at the center from the value of the first maximum. The evolution of two initial data with $\omega/k=4.6680$ is shown, corresponding to different resolutions in the spectral code, one with $N_t=17$ and the other with $33$ temporal coefficients.  \label{expincfig}}
\end{figure}

The instability does not mean that the configuration decays, since energy cannot disperse to infinity as in the asymptotically flat case. The AdS background acts as an effective bounding box. The instability is quite weak, and appears very slowly. Initially the field oscillates periodically, with the predicted frequency. However, slowly the solution becomes more and more different from the simple periodic state. This can be seen most easily by showing the upper and lower envelope of the oscillations. On Fig.~\ref{minmaxfig} each local maximum and minimum of the oscillation of the field at the center is represented by a dot. The initial data corresponds to $\omega/k=4.6680$ on the lower branch of the first panel of Fig.~\ref{orik}.
For a very long initial period, every second maximum gets slightly bigger, while the others smaller. This goes on until the change becomes comparable to the original amplitude, then the configuration tends to be more periodic for some time, but later the difference from periodicity increases again. The difference from periodicity increases exponentially for a long initial period. This can be seen on Fig.~\ref{expincfig}, where the absolute value of the difference of each maximum from the value of the first maximum is plotted as a function of time.
The exponential increase indicates the existence of a single unstable mode. It can also be seen on the figure, that the more precisely the initial data is calculated, the later the deviation from periodicity appears. This exponentially increasing deviation from periodicity can be observed at all the unstable states that we have calculated by the spectral code and evolved by the time evolution code. For large amplitude configurations, the later stage, where the error is not monotonically increasing anymore, is not as simple as for the $\omega/k=4.6680$ case, the evolution appears quite noisy, without the periodically appearing structures that can be seen on Fig.~\ref{minmaxfig}.

The stability properties of the $k^2=0.001$ states that can be seen on the second panel of Fig.~\ref{orik} are somewhat more complicated. Most of the configurations on this figure are evolving stably periodically, but there are unstable states close to the resonance-like structures, and also for large amplitudes. At the first small resonance, which can be seen on the first panel of Fig.~\ref{reson_std}, there are unstable states at the end of the downwards pointing branch for $32.523\leq\omega/k\leq 32.523224$. On the upper branch of the resonance shown on the first panel of Fig.~\ref{reson_2} configurations are unstable in the short interval at the end of the line for $31.453\leq\omega/k\leq 31.455$. States on the lower branch of this resonance are unstable for $31.428\leq\omega/k\leq 31.495$. There are unstable states also for  $\omega/k\leq 31.1$. The question of stability is obviously very complicated when $k$ is small, and it would certainly deserve further studies in the future, together with the better understanding of the resonance-like structures appearing at certain frequencies.

\section{Small-amplitude expansion} \label{smallamplsec}

For general $U(\phi)$ potentials small amplitude localized configurations on AdS can be expected to behave similarly to the Klein-Gordon breathers discussed earlier. In this section we carry out an expansion procedure for self-interacting scalar breathers in terms of an amplitude parameter, and in the end we compare the results to numerical results in order to see the range of validity of this approximation. We use the radial coordinate $z$ defined in \eqref{zcoord}, which is particularly useful when dealing with the hypergeometric functions appearing in the problem. In order to factor out the asymptotic behavior of the scalar field we use the rescaled field variable $\psi$ defined in \eqref{psidef}. For small-amplitude configurations the field probe the potential close to its minimum, so we use the expansion \eqref{phiexp}, and characterize the potential in terms of the scalar field mass $m$ and the expansion coefficients $g_k$. Taking out the mass term from the potential, and denoting the sum of the other
terms
by $\tilde U(\phi)$ according to \eqref{utildedef}, the field equation can be written into the form
\begin{equation}
-\partial_{t}^2\psi+4k^2D_z\psi-k^2\lambda_+^2\psi
=\frac{1}{z^{\lambda_+/2+1}}\tilde U'(z^{\lambda_+/2}\psi) \ , \label{zpsieq}
\end{equation}
where $D_z$ is the differential operator defined in \eqref{dzdef}.

We look for periodic solutions with basic frequency $\omega$, so we Fourier decompose the scalar field as
\begin{equation}
\psi=\sum_{j=0}^\infty\psi_j\cos(j\omega t) \ .
\end{equation}
We further expand the Fourier components of the scalar field in powers of a small parameter $\varepsilon$
\begin{equation}
\psi_j=\sum_{n=1}^\infty\psi_{j}^{(n)}\varepsilon^n \ .
\end{equation}
We will build up the formalism in such a way that $\varepsilon$ gives the amplitude of the $\cos(\omega t)$ Fourier mode at the center. The frequency $\omega$ may also depend on the small parameter $\varepsilon$, which we take into account by expanding the square of $\omega$ as:
\begin{equation}
\omega^2=\omega_0^2\left(1+\sum_{j=1}^\infty\omega_j\varepsilon^j\right) \ , \label{omegaexpeq}
\end{equation}
where $\omega_0$ and $\omega_j$ are constants.

Substituting into the field equation \eqref{zpsieq}, to leading $\varepsilon$ order we obtain
\begin{equation}
D_z\psi_j^{(1)}-\frac{1}{4}\left(\lambda_+^2-\frac{j^2\omega_0^2}{k^2}
\right)\psi_j^{(1)}=0 \ .  \label{psi1jeq}
\end{equation}
After replacing $j\omega_0$ by $\omega$ this is exactly the hypergeometric differential equation \eqref{dzqeq}, which we have studied in detail earlier. We have seen that this equation has solutions that are regular both at the center and at infinity only if
\begin{equation}
\frac{j\omega_0}{k}=\lambda_++2n ,
\end{equation}
where $n$ is a non-negative integer. The corresponding solution has $n$ nodes, and  in the nodeless case $\psi_j^{(1)}$ is just a constant function. We are interested now in solutions which tend to the nodeless Klein-Gordon breather in the small-amplitude limit, so we take
\begin{equation}
\psi_1^{(1)}=1 \  \mathrm{and} \ \psi_j^{(1)}=0 \ \mathrm{for} \ j\neq 1 \ .
\end{equation}
This choice fixes the leading order behavior of the frequency,
\begin{equation}
\omega_0=k\lambda_+ \ . \label{omega0eq}
\end{equation}
The normalization $\psi_1^{(1)}=1$ ensures that to leading order $\varepsilon$ yields the  amplitude of the rescaled scalar field $\psi$. Going to higher orders in $\varepsilon$, we will see that the equations determining $\psi_1^{(n)}$ will only contain the derivatives of $\psi_1^{(n)}$. Hence to any solution we can add an arbitrary constant. We use this freedom to set $\psi_1^{(n)}=0$ at the center $z=1$ for $n>1$. By this choice we ensure that the central value of the first Fourier mode is exactly $\varepsilon$,
\begin{equation}
\psi_1^\textrm{cent}=\varepsilon \ .
\end{equation}
We note that the same is true for the physical field $\phi=z^{\lambda_+/2}\psi$, since the two fields agree at the center.

To $\varepsilon^2$ order, the Fourier components of \eqref{zpsieq} yield
\begin{align}
D_z\psi_0^{(2)}-\frac{1}{4}\lambda_+^2\psi_0^{(2)}
&=\frac{g_2}{8k^2} z^{\lambda_+/2-1} \ , \label{phi20eq}\\
D_z\psi_1^{(2)}&=-\frac{\omega_1}{4}\lambda_+^2 \ , \label{phi21eq} \\
D_z\psi_2^{(2)}+\frac{3}{4}\lambda_+^2\psi_2^{(2)}
&=\frac{g_2}{8k^2} z^{\lambda_+/2-1} \ , \label{phi22eq} \\
D_z\psi_j^{(2)}+\frac{j^2-1}{4}\lambda_+^2\psi_j^{(2)}
&=0 \ \ \mathrm{for} \ \ j\geq3 \ . \label{phi2jeq}
\end{align}
The homogeneous terms on the left hand sides form the same hypergeometric equation as in \eqref{psi1jeq}, except that the already known value of $\omega_0$ has been substituted from \eqref{omega0eq}. This also means that the constants in \eqref{abceq} are now
\begin{equation}
a=\frac{1+j}{2}\lambda_+ \ , \quad b=\frac{1-j}{2}\lambda_+ \ .
\end{equation}

For $j\geq3$ the equations have no inhomogeneous source terms. Equation \eqref{phi2jeq} has regular localized solution only if $j\omega_0=k(\lambda_++2n)$ for some nonnegative integer $n$, which is satisfied if $(j-1)\lambda_+=2n$. Since for nonzero scalar field mass $m$ generally $\lambda_+$ is not a rational number, this can hold only for very exceptional cases.
Even if there is a localized solution for $j\geq3$, we take it with amplitude $0$. Based on our experience with massless fields we are confident that at higher orders in the $\varepsilon$ expansion
new conditions arise forcing the vanishing of these terms. From now on we assume that $\psi_j^{(2)}=0$ for $j\geq3$.

The only solution of equation \eqref{phi22eq} which is regular at both the origin and infinity is
\begin{equation}
\psi_2^{(2)}=\frac{g_2}{2k^2\lambda_+(3\lambda_+-D)}z^{\lambda_+/2} \ .
\end{equation}

At all orders of the expansion we have to solve inhomogeneous differential equations of the form
\begin{equation}
D_z\psi_j^{(i)}+\frac{j^2-1}{4}\lambda_+^2\psi_j^{(i)}=h \ ,
\end{equation}
where $h$ is a given function of $z$. Two pairs of solutions for the homogeneous part, $w_1$, $w_2$, $w_3$ and $w_4$, are given in \eqref{homsolz0} and \eqref{homsolz1}. The general  solution of the inhomogeneous equation can be written in terms of any two independent homogeneous solution as
\begin{equation}
\psi_j^{(i)}=w_b\int_{z_1}^z\frac{w_a}{W_{ab}}\tilde hdz
-w_a\int_{z_2}^z\frac{w_b}{W_{ab}}\tilde hdz \ ,  \label{inhomij}
\end{equation}
where $z_1$ and $z_2$ are arbitrary constants,
\begin{equation}
\tilde h=\frac{h}{z(1-z)} \ , \label{htilde}
\end{equation}
and the Wronskian is $W_{ab}=w_a d_z w_b-w_b d_z w_a$. We need to use two different choices for $a$ and $b$,
\begin{align}
W_{12}&=(1-c)z^{-c}(1-z)^{c-a-b-1}
=-\frac{\lambda}{z^{\lambda+1}(1-z)^{D/2}} \ , \\
W_{13}&=\frac{\Gamma(a+b+1-c)\Gamma(c-1)}{\Gamma(a)\Gamma(b)}W_{12}
=-\frac{\lambda\Gamma(\lambda)\Gamma(D/2)}
{\Gamma\left(\frac{1+j}{2}\lambda_+\right)
\Gamma\left(\frac{1-j}{2}\lambda_+\right)
z^{\lambda+1}(1-z)^{D/2}} \ ,  \label{w13eq}
\end{align}
where we have introduced the notation
\begin{equation}
\lambda=\lambda_+-\frac{D}{2}=\sqrt{\frac{D^2}{4}+\frac{m^2}{k^2}} \ .
\end{equation}
The solution $w_1$ is always regular at $z=0$, but generally singular at $z=1$. On the other hand, $w_3$ is always regular at $z=1$ but generally singular at $z=0$. Hence if $w_1$ and $w_3$ are linearly independent, it is most natural to use them in the expression \eqref{inhomij} of the inhomogeneous solution. We are interested in the inhomogeneous solution which is regular both at the center and infinity. This fixes the values of the integration limits $z_1$ and $z_2$,
\begin{equation}
\psi_j^{(i)}=w_3\int_{0}^z\frac{w_1}{W_{13}}\tilde hdz
-w_1\int_{1}^z\frac{w_3}{W_{13}}\tilde hdz \ .  \label{inhom13}
\end{equation}
With this choice the singularity of $w_3$ at $z=0$ is compensated by the vanishing of the first integral, and similarly, the singularity of $w_1$ at $z=1$ is compensated by the second integral. However, the expression \eqref{inhom13} for the inhomogeneous solution is only valid if $w_1$ and $w_3$ is linearly independent, which is always true if $W_{13}$ is nonzero. According to \eqref{w13eq}, $W_{13}$ can only be zero if $\frac{1-j}{2}\lambda_+$ is a nonpositive integer. Certainly this is the case if $j=1$, so we have to treat the $\cos(\omega t)$ Fourier component equations for $\psi_1^{(i)}$ separately at each order. However in general, for nonzero scalar field mass $\lambda_+$ is not a rational number, and consequently for the $j\neq1$ equations $W_{13}$ is not zero.

For the $j=1$ case $a=\lambda_+$ and $b=0$, consequently $w_1=w_3=1$. The solution $w_2$ diverges both at $z=0$ and $z=1$. In this case we have to use the pair of solutions $w_1=1$ and $w_2$ to generate the inhomogeneous solution,
\begin{equation}
\psi_1^{(i)}=w_2\int_{0}^z\frac{1}{W_{12}}\tilde hdz
-\int_{1}^z\frac{w_2}{W_{12}}\tilde hdz \ .  \label{inhom12}
\end{equation}
The singularity of $w_2$ at $z=0$ is already compensated by the choice of the lower limit in the first integral, but in order to cancel the divergence of $w_2$ at $z=1$ we need to require that
\begin{equation}
\int_0^1\frac{1}{W_{12}}\tilde hdz=0 \ .  \label{int01}
\end{equation}
The lower limit in the second integral of \eqref{inhom12} was set to $1$ in order to set the central value of $\psi_1^{(i)}$ to $0$. At $z=1$ the first term in \eqref{inhom12} also tends to zero, which can be seen in the following way. According to \eqref{homsolz1}, the homogeneous solution $w_2$ diverges as $(1-z)^{1-D/2}$, the same way as $w_4$. Since $h$ is not singular at $z=1$, the integrand tends to zero as $(1-z)^{D/2-1}$. Since the whole integral from $0$ to $1$ vanishes, the integral from $0$ to $z$ can be replaced by integral from $1$ to $z$. Consequently, the integral tends to zero as $(1-z)^{D/2}$, and the whole first term as $1-z$.

According to \eqref{phi21eq}, for $\psi_1^{(2)}$ we have $h=-\omega_1\lambda_+^2/4$, and the condition \eqref{int01} can be written as
\begin{equation}
0=\int_0^1\frac{\omega_1\lambda_+^2}{4\lambda}z^\lambda(1-z)^{D/2-1}dz
=\frac{\lambda_+^2\Gamma(\lambda)\Gamma(D/2)}{4\Gamma(\lambda_++1)}\omega_1 \ .
\end{equation}
From this it follows that necessarily $\omega_1=0$, and then we can take $\psi_1^{(2)}=0$. Continuing to higher orders in the $\varepsilon$ expansion, it turns out similarly that $\omega_i=0$ for all odd $i$.

For the $j=0$ case the solution of \eqref{phi20eq} for $\psi_0^{(2)}$ can be obtained using \eqref{inhom13}, where $h=g_2z^{\lambda_+/2-1}/(8k^2)$. For zero mass scalar fields $\lambda_+=D$, and the solution can be calculated explicitly. For specific choices of spatial dimension $D$, in the $m=0$ case we obtain:
\begin{align}
&D=1 \ \ : \quad \psi_0^{(2)}=-\frac{g_2}{16k^2\cos x}(\pi^2-4x^2) \ , \\
&D=2 \ \ : \quad \psi_0^{(2)}=-\frac{g_2}{8k^2} \ , \\
&D=3 \ \ : \quad \psi_0^{(2)}=-\frac{g_2}{512k^2\sin x\cos^3 x}
\left[32x\cos x+(6-4\pi^2+16x^2)\sin x+7\sin(3x)+\sin(5x)\right] \ , \\
&D=4 \ \ : \quad \psi_0^{(2)}=-\frac{g_2}{288k^2}\left[5+2\cos(2x)\right] \ .
\end{align}

\begin{figure}[!hbp]
\includegraphics[width=10cm]{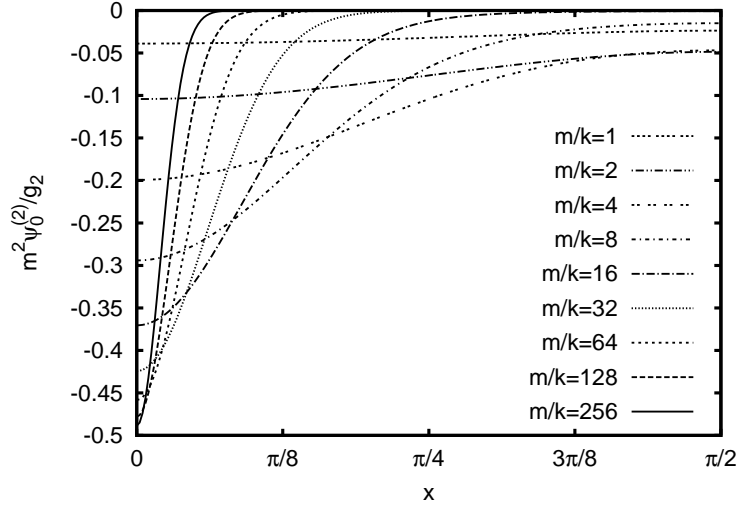}
\caption{The radial dependence of $m^2\psi_0^{(2)}/g_2$ for various $m/k$ values. \label{figpsi20x}}
\end{figure}

For nonzero scalar field mass the integrals in \eqref{inhom13} determining $\psi_0^{(2)}$ can either be calculated numerically, or alternatively, the result can be written into a complicated expression involving ${}_3F_2$ hypergeometric functions. In order to do that, we have to express $w_3$ in the second integral by $w_1$ and $w_2$, using \eqref{w3trans}. Then we have to evaluate integrals of expressions like $z^\alpha(1-z)^{D/2-1}{}_2F_1(a,b;c;z)$, where $\alpha$ is some constant. In order to get rid of the $(1-z)^{D/2-1}$ factor we can apply the identity\cite{dlmf}
\begin{equation}
{}_2F_1(a,b;c;z)=(1-z)^{c-a-b}{}_2F_1(c-a,c-b;c;z) \ .
\end{equation}
Then the integrals can be performed using
\begin{equation}
\int z^\alpha{}_2F_1(a,b;c;z)dz=\frac{z^{\alpha+1}}{\alpha+1}
\ {}_3F_2(a,b,\alpha+1;c,\alpha+2;z) \ .
\end{equation}
We do not give here the resulting expression for $\psi_0^{(2)}$ because of its complexity. For given $D$ the left hand side of \eqref{phi20eq} depends on the constants only through the ratio $m/k$. Consequently, $m^2\psi_0^{(2)}/g_2$ will depend only on $m/k$. In case of $D=3$, for specific values of $m/k$ the radial dependence of $m^2\psi_0^{(2)}/g_2$ is plotted on Fig.~\ref{figpsi20x}.
We note that for the $\phi^4$ potential of the form \eqref{phi4pot} we have $g_2/m^2=-3/2$. It helps the numerical evaluation to notice that close to $z=0$ the first term becomes negligible in \eqref{inhom13}, while at $z=1$ the second term tends to zero. At the center the value of $m^2\psi_0^{(2)}/g_2$ decreases monotonically with growing $m/k$, approaching the limit value $-1/2$. The value of $m^2\psi_0^{(2)}/g_2$ at infinity $x=\pi/2$ has a minimum of $-0.0534448$ at $m/k=2.77615$, and tends to zero when $m/k$ tends to zero or infinity.

The Fourier components of the field equation \eqref{zpsieq} to $\varepsilon^3$ order give
\begin{align}
D_z\psi_j^{(3)}+\frac{j^2-1}{4}\lambda_+^2\psi_j^{(3)}
&=0 \ \ \mathrm{for} \ \ j=0, \ j=2 \ \mathrm{and} \  j\geq4 \ , \\
D_z\psi_1^{(3)}
&=-\frac{\omega_2}{4}\lambda_+^2+\frac{g_2}{4k^2} z^{\lambda_+/2-1}
\left(2\psi_0^{(2)}+\psi_2^{(2)}\right)
+\frac{3g_3}{16k^2}z^{\lambda_+-1} \ , \label{eqpsi31}\\
D_z\psi_3^{(3)}+2\lambda_+^2\psi_3^{(3)}
&=\frac{g_2}{4k^2} z^{\lambda_+/2-1}\psi_2^{(2)}
+\frac{g_3}{16k^2}z^{\lambda_+-1} \ . \label{eqpsi33}
\end{align}
Except for $j=1$ and $j=3$ we can take $\psi_j^{(3)}=0$. The regular solution of \eqref{eqpsi33} can be given explicitly,
\begin{equation}
\psi_3^{(3)}=\frac{2g_2^2+g_3k^2\lambda_+(3\lambda_+-D)}
{8k^4\lambda_+^2(3\lambda_+-D)(4\lambda_+-D)}z^{\lambda_+} \ .
\end{equation}

\begin{figure}[!hbp]
\includegraphics[width=10cm]{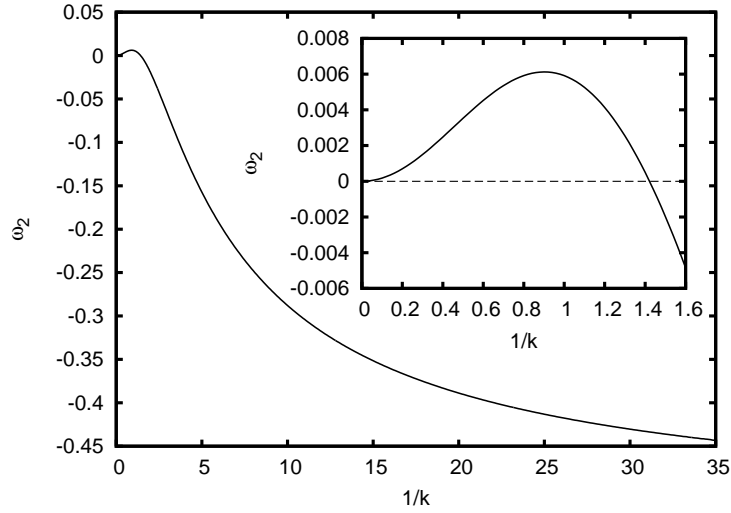}
\caption{Value of $\omega_2$ for the symmetric $\phi^4$ potential as a function of $1/k$. \label{figom2}}
\end{figure}

Equation \eqref{eqpsi31} has localized regular solutions for $\psi_1^{(3)}$ only for specific values of $\omega_2$. Similarly to the $\psi_1^{(2)}$ equation, the solutions of the homogeneous part can be written in terms of $w_1=1$ and $w_2$ according to \eqref{inhom12}. Defining $h$ as the right hand side of \eqref{eqpsi31}, the condition of the existence of finite regular solutions is given by \eqref{int01}. For $m=0$ the integral can be calculated explicitly. The result for $D=3$ spatial dimensions is
\begin{equation}
\omega_2=\frac{1}{48}\left[\left(\frac{\pi^2}{9}
-\frac{301}{216}\right)\frac{g_2^2}{k^4}+\frac{7g_3}{4k^2}\right]  .
\end{equation}
In general, for nonzero $m$, because of the complicated form of $\psi_0^{(2)}$, the integral in \eqref{int01} can be evaluated only numerically. For the $\phi^4$ potential given by \eqref{phi4pot} the dependence of $\omega_2$ on $k$ is shown on Fig.~\ref{figom2}.
The value of $\omega_2$ has a maximum at $1/k=0.9020$ with value $0.006123$, and changes sign at $1/k=1.4216$. The importance in the sign of $\omega_2$ lies in the fact that, according to \eqref{omegaexpeq}, if $\omega_2>0$, i.e.~when $k<0.70343$, the frequency of small amplitude configurations is larger than the frequency $\omega_0$ of the same mass Klein-Gordon breather, and increases with increasing amplitude. The opposite is true for $k>0.70343$, when $\omega_2$ is negative. The numerical results show that this tendency of how the frequency changes continues to quite large amplitude states as well.

After setting $\omega_2$ appropriately, the function $\psi_1^{(3)}$ can be calculated using \eqref{inhom12}. For the $m=0$ case the integrals in \eqref{inhom12} can be calculated, yielding a long expression involving polylogarithm functions that we do not show here.

It is worthwhile to go to higher orders in the $\varepsilon$ expansion, since it increases very much the range of validity of the approximation. However, because of the complexity of the equations we only show the results for $D=3$ spatial dimensions and for the $\phi^4$ potential given in \eqref{phi4pot}. In this case we will also compare to the numerical results. At $\varepsilon^4$ order we get the following two differential equations, which can be solved by a numerical method
\begin{align}
 D_z\psi_0^{(4)}-\frac{1}{4}\lambda_+^2\psi_0^{(4)}
=&-\frac{3}{8k^2} z^{\lambda_+/2-1}\left[\left(\psi_0^{(2)}\right)^2+\psi_1^{(3)}\right]
+\frac{3}{16k^2} z^{\lambda_+-1}\psi_0^{(2)}
-\frac{3z^{3\lambda_+/2-1}\left[1+2k^2\lambda_+(\lambda_+-1)\right]}
{256k^6\lambda_+^2(\lambda_+-1)^2}
\ , \label{phi40eq}\\
D_z\psi_2^{(4)}+\frac{3}{4}\lambda_+^2\psi_2^{(4)}
=&\frac{3z^{\lambda_+-1}\left[1+k^2\lambda_+(\lambda_+-1)\right]}
{16k^4\lambda_+(\lambda_+-1)}\psi_0^{(2)}
-\frac{3}{8k^2} z^{\lambda_+/2-1}\psi_1^{(3)}
+\frac{\omega_2\lambda_+z^{\lambda_+/2}}{4k^2(\lambda_+-1)} \notag\\
&-\frac{3z^{3\lambda_+/2-1}\left[3+k^2\lambda_+(9\lambda_+-7)\right]}
{128k^6\lambda_+^2(\lambda_+-1)(4\lambda_+-3)}
\ . \label{phi42eq}
\end{align}
A third equation can be solved explicitly, giving
\begin{equation}
\psi_4^{(4)}=\frac{9-10\lambda_+-2k^2\lambda_+(\lambda_+-1)(5\lambda_+-4)}
{64k^6\lambda_+^3(\lambda_+-1)^2(4\lambda_+-3)(5\lambda_+-3)}z^{3\lambda_+/2} \ .
\end{equation}
At $\varepsilon^5$ order the important equation is
\begin{align}
D_z\psi_1^{(5)}
=&-\frac{\lambda_+^2}{4}\left(\omega_4+\omega_2\psi_1^{(3)}\right)
-\frac{3}{8k^2} z^{\lambda_+/2-1}
\left(2\psi_0^{(4)}+\psi_2^{(4)}+2\psi_1^{(3)}\psi_0^{(2)}\right)
+\frac{3}{8k^2}z^{\lambda_+-1}\left(\psi_0^{(2)}\right)^2 \notag\\
&+\frac{3z^{\lambda_+-1}\left[1+3k^2\lambda_+(\lambda_+-1)\right]}
{32k^4\lambda_+(\lambda_+-1)}\psi_1^{(3)}
+\frac{3z^{2\lambda_+-1}\left[3+k^2\lambda_+(6\lambda_+-5)+k^4\lambda_+^2(\lambda_+-1)^2\right]}
{512k^8\lambda_+^3(\lambda_+-1)^2(4\lambda_+-3)} \ ,
\end{align}
since it gives the condition determining $\omega_4$. These equations can be solved by calculating integrals using \eqref{inhom13} and \eqref{inhom12}. However, technically it is easier to solve the differential equations by a Runge-Kutta shooting method starting from the center. It is also advantageous to use the radial coordinate $x$ in this case, since then the derivatives of the functions at the center $x=0$ vanish.

\begin{figure}[!htbp]
\includegraphics[width=10cm]{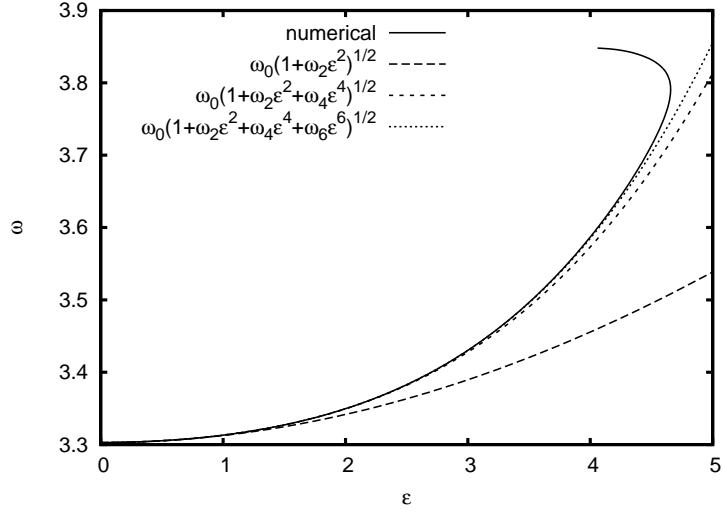}
\caption{The comparison of the numerically calculated frequency to the first few order results of the small-amplitude expansion for $k=1$. The parameter $\varepsilon$ gives the central value of the first Fourier mode. \label{omepfig}}
\end{figure}

For the rest of this subsection we set $k=1$, and compare the small-amplitude expansion results to the numerical data obtained by the spectral code in Sec.~\ref{persolsec}. The dependence of the frequency on the parameter $\epsilon$, which is just the central value of the first Fourier mode, can be seen on Fig.~\ref{omepfig}.
In the $k=1$ case $\omega_0=(3+\sqrt{13})/2\approx3.3027756$, and from the $\varepsilon$ expansion we get $\omega_2=0.00591653$ and $\omega_4=0.00029594$. The value $\omega_6=1.90\cdot10^{-6}$ comes from a fit to the numerical data, since we did not go to high enough order to get this coefficient from the expansion. The results indicate that the small-amplitude expansion is most likely convergent for as large values of the expansion parameter as $\varepsilon=4$. However, for frequencies above $\omega=3.8$ the central amplitude $\varepsilon$ is not an increasing function of the frequency $\omega$, and the expansion is not valid anymore.

\begin{figure}[!htbp]
\includegraphics[width=10cm]{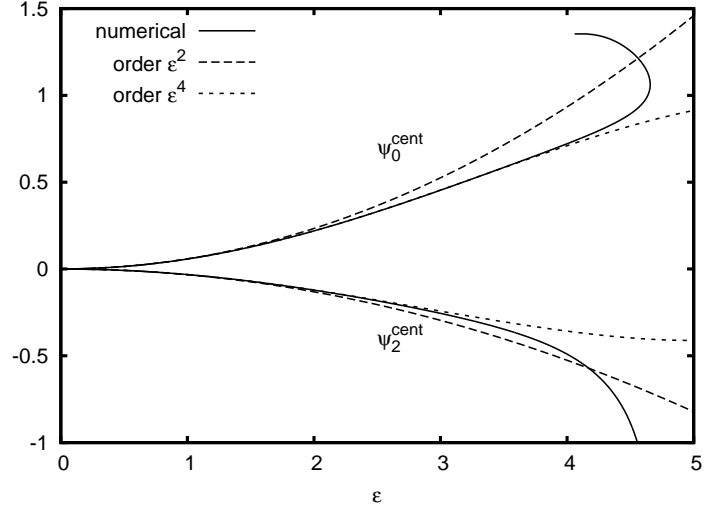}
\caption{The central value of the zeroth and second modes. \label{phi02cfig}}
\end{figure}

The central values of the zeroth and second Fourier mode can be seen on Fig.~\ref{phi02cfig}, as functions of $\varepsilon$.
In this case, from the small-amplitude expansion we get $\psi_0^\mathrm{cent}=0.058351\varepsilon^2-0.0008732\varepsilon^4$, and $\psi_2^\mathrm{cent}=-0.032871\varepsilon^2+0.0006558\varepsilon^4$.

\begin{figure}[!htbp]
\includegraphics[width=10cm]{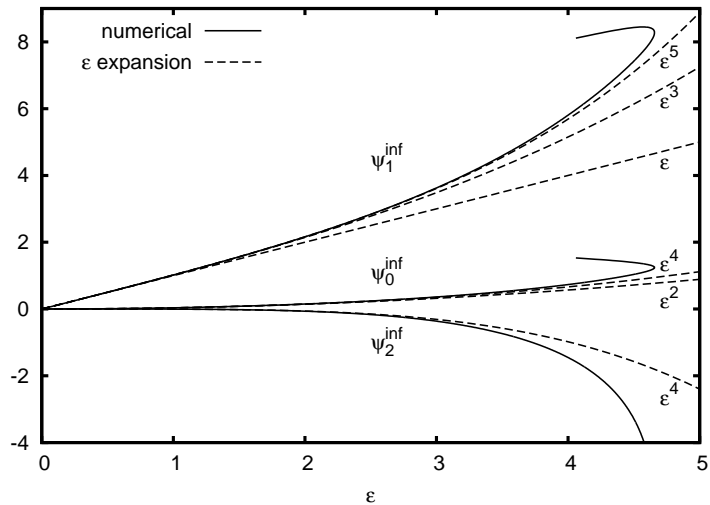}
\caption{The values of the first three Fourier modes at infinity. \label{phi012ifig}}
\end{figure}

On Fig.~\ref{phi012ifig} the value of the first three Fourier modes of $\psi$ are given at infinity $x=\pi/2$.
Up to order $\varepsilon^5$ the small-amplitude results are $\psi_0^\mathrm{inf}=0.035537\varepsilon^2+0.0003675\varepsilon^4$, $\psi_1^\mathrm{inf}=\varepsilon+0.017988\varepsilon^3+0.0005220\varepsilon^5$ and $\psi_2^\mathrm{inf}=-0.003837\varepsilon^4$. The very good agreement of the results obtained by the different methods ensures us that the boundary conditions at infinity are really treated very precisely by the spectral numerical code in Sec.~\ref{persolsec}.

\section{Expansion for small cosmological constant} \label{smalksec}

In this section we apply an expansion procedure, which had been first worked out on a flat background in \cite{dashen,SK,Kichenassamy,FFHL2} for small-amplitude  oscillons formed by self-interacting scalar fields\cite{BogMak2,Gleiser,CopelGM95}, and later also for oscillons on an expanding de Sitter background, when the cosmological constant is small\cite{fggirs}. A similar expansion for oscillatons\cite{Seidel1,Seidel2}, formed by a self-gravitating scalar field, was also worked out in \cite{oscillatons}. We use Schwarzschild area coordinates \eqref{schcoord}, where the field equation for the scalar takes the form \eqref{fieldeq}. Since oscillons on a flat background only exist for positive mass scalar fields, the procedure in this section is valid only for nonzero scalar field mass $m$. In order to simplify the equations, in this case we use the rescaled expansion coefficients
\begin{equation}
\tilde g_j=\frac{g_j}{m^2}
\end{equation}
instead of the $g_j$ to characterize the scalar potential. With this notation it is possible to take out a common $m^2$ factor from the expansion \eqref{phiexp} of $U(\phi)$.

We expand the scalar field in powers of a small parameter $\varepsilon$
\begin{equation}
\phi=\sum_{n=1}^\infty\phi^{(n)}\varepsilon^n \ . \label{phiepsexp}
\end{equation}

\subsection{Coordinates and parameters}

Since on Minkowski background the size of small amplitude oscillons scales as $1/\varepsilon$, we use a rescaled radial coordinate
\begin{equation}
\rho=\varepsilon m r \ , \label{rhortrans}
\end{equation}
which makes spatial derivatives one order smaller. For the small-amplitude expansion it is advantageous to use a time coordinate $\tau$ in which the (angular) frequency of the basic oscillation mode of
$\phi$ is one,
\begin{equation}
\tau=m\tilde\omega t \ . \label{taumtot}
\end{equation}
We note that the time coordinate $\tau$ defined here is different from the coordinate $\tau$ used in the conformal form of the metric \eqref{confcord}. In this section we only work in Schwarzschild area coordinates, and only use the $\tau$ defined in \eqref{taumtot}. The physical frequency of the oscillations is then $\omega=m\tilde\omega$. In general, the frequency  will depend on the amplitude of the oscillations, so $\omega$ will be $\varepsilon$ dependent. Since in the zero amplitude limit $\omega$ approaches $m$, we represent this dependence by the following expansion
\begin{equation}
\tilde\omega^2=1+\sum_{n=1}^\infty\omega_{2n}\varepsilon^{2n} \ ,  \label{omegatexp}
\end{equation}
where $\omega_k$ are constants. One could include odd powers of $\varepsilon$ into the expansion of $\tilde\omega^2$, but the coefficients of those terms turn out to be zero if we require that the solution remains bounded as time passes\cite{FFHL2}. In general, there is a parametrization freedom on how one assigns the small parameter $\varepsilon$ to the different physical states. On Minkowski background it is advantageous to relate $\varepsilon$ to the frequency by setting $\tilde\omega=\sqrt{1-\varepsilon^2}$. On AdS background $\tilde\omega$ may be both larger and smaller than one, hence a different approach will be chosen.

We introduce a rescaled $k$ cosmological parameter $\kappa$ by
\begin{equation}
k=\varepsilon^2m\kappa  \ ,   \label{kepmkappa}
\end{equation}
and assume that $\kappa$ is some fixed $\mathcal{O}(1)$ constant. This means that $k$ tends to zero in the small $\varepsilon$ limit. Our expansion in this section is a simultaneous small-amplitude and small $k$ expansion with one common small parameter $\varepsilon$. The expansion procedure is valid if $k/m$ is small. The choice \eqref{kepmkappa} ensures that the oscillon size, which is approximately $1/\varepsilon$, remains small compared to the curvature scale $1/k$ even in the $\varepsilon\to 0$ limit.

\subsection{Expansion procedure}

Substituting the expansion \eqref{phiepsexp} into the field equation \eqref{fieldeq}, the leading $\varepsilon$ order terms give
\begin{equation}
\phi^{(1)}_{,\tau\tau}+\phi^{(1)}=0 \ ,
\end{equation}
which has the general solution $\phi^{(1)}=p_1\cos(\tau+\delta)$, where $p_1$ and $\delta$ are some functions of $\rho$. It can be shown, that continuing to higher orders in $\varepsilon$, the requirement that $\phi^{(3)}$ remains bounded fixes $\delta$ to be a constant, which can be set to zero\cite{FFHL2}. To make the equations as short as possible we set $\delta=0$ here, so we have
\begin{equation}
\phi^{(1)}=p_1\cos\tau \ .  \label{phi1eq}
\end{equation}

The $\varepsilon^2$ order component of \eqref{fieldeq} gives an equation which determines the time dependence of $\phi^{(2)}$, resulting in
\begin{equation}
\phi^{(2)}=\frac{\tilde g_2}{6}p_1^2\left[\cos(2\tau)-3\right] . \label{phi2eq}
\end{equation}
It would be possible to include two further terms, $p_2\cos\tau$ and $q_2\sin\tau$ in \eqref{phi2eq}, where $p_2$ and $q_2$ are some functions of $\rho$. However, $q_2$ can be set to zero by a small radius dependent shift in the time coordinate. Continuing to higher orders, one can show that there are no $\sin(n\tau)$ terms in the expansion at all, which is a consequence of the time reflection symmetry of the system. At $\varepsilon^4$ order one would get a differential equation for $p_2$, which has no regular localized solution. In general, it can be checked that there are no $\cos\tau$ terms in $\phi^{(2n)}$. A new radial function, $p_{2n+1}$, arises at every odd order $\epsilon^{2n+1}$ for $n\geq0$ integer, and a differential equation determining it comes at order $\epsilon^{2n+3}$.

The $\varepsilon^3$ order field equation gives the time dependence of $\phi^{(3)}$, but there are terms which blow up in time. The vanishing of the terms proportional to $\tau\sin\tau$ in $\phi^{(3)}$ requires that
\begin{equation}
p_{1,\rho\rho}+\frac{D-1}{\rho}p_{1,\rho}+\left(\omega_2-\kappa^2\rho^2\right)p_1
+\gamma p_1^3=0 \ , \label{p1eq}
\end{equation}
where
\begin{equation}
\gamma=\frac{5}{6}\tilde g_2^2-\frac{3}{4}\tilde g_3 \ .
\end{equation}
The spatially localized solution of equation \eqref{p1eq} yields the spatial profile of the oscillon to leading order, since $\phi=\varepsilon p_1\cos\tau+\mathcal{O}(\varepsilon^2)$. The time dependence of $\phi^{(3)}$ is then given by
\begin{equation}
\phi^{(3)}=p_3\cos\tau+\frac{1}{96}p_1^3(2\tilde g_2^2+3\tilde g_3)\cos(3\tau) \ , \label{phi3eq}
\end{equation}
where $p_3$ is a function of $\rho$.

The $\varepsilon^4$ component of \eqref{fieldeq} gives the time dependence of $\phi^{(4)}$,
\begin{align}
\phi^{(4)}=&\frac{\tilde g_2}{3}p_3p_1[\cos(2\tau)-3]
-\frac{\tilde g_2}{9}\left(p_{1,\rho}\right)^2[\cos(2\tau)+9]
+\frac{\tilde g_2}{9}p_1^2(\rho^2\kappa^2-\omega_2)[\cos(2\tau)-9]  \notag\\
&+\left[\frac{\tilde g_2}{72}(\tilde g_2^2-\gamma)+\frac{\tilde g_4}{120}\right]p_1^4\cos(4\tau)
+\left[-\frac{\tilde g_2}{72}(16\tilde g_2^2-23\gamma)+\frac{\tilde g_4}{6}\right]p_1^4\cos(2\tau) \label{phi4eq}\\
&+\left[\frac{\tilde g_2}{72}(31\tilde g_2^2+12\gamma)-\frac{3\tilde g_4}{8}\right]p_1^4 \ , \notag
\end{align}
without any new function arising. The function $p_3$ is determined by requiring the vanishing of the $\tau\cos\tau$ terms in the $\varepsilon^5$ order field equation,
\begin{align}
&p_{3,\rho\rho}+\frac{D-1}{\rho}p_{3,\rho}
+\left(\omega_2-\kappa^2\rho^2+3\gamma p_1^2\right)p_3
+\frac{19}{9}\tilde g_2^2p_1\left(p_{1,\rho}\right)^2
+2\kappa^2\rho p_{1,\rho} \notag\\
&+\left(-\frac{35}{27}\tilde g_2^4+\frac{\tilde g_2^2}{6}\gamma-\frac{\gamma^2}{24}
+\frac{7}{4}\tilde g_2\tilde g_4-\frac{5}{8}\tilde g_5\right)p_1^5
+\left[\frac{17}{9}\tilde g_2^2(\kappa^2\rho^2-\omega_2)-\kappa^2\gamma\rho^2\right]p_1^3 \notag\\
&+\left[\omega_4+2\kappa^2\rho^2(\kappa^2\rho^2-\omega_2)\right]p_1
=0 \ . \label{p3eq}
\end{align}
After numerically obtaining the localized solution $p_1$ of \eqref{p1eq} for a specific choice of $\omega_2$, we can substitute it into \eqref{p3eq} and solve for $p_3$. However, $\omega_4$ is not specified yet, and for each particular choice of $\omega_4$, equation \eqref{p3eq} has a unique localized solution with a regular center. We fix the freedom in $\omega_4$ by requiring that $p_3$ has to be zero at the center $\rho=0$. By this we choose a particular way of how we parametrize the various configurations by $\omega_2$. Continuing to higher orders, we can always require that $p_{2n+1}$ vanishes at the center. Then the differential equation for $p_{2n+1}$ has localized solution only for a certain value of $\omega_{2n+2}$. It can be checked that for $n\geq 2$ all $\cos\tau$ terms in $\phi^{(n)}$ will have a multiplying factor that vanishes at the center. Consequently, the central value of the first Fourier mode of $\phi$ agrees to all orders with $\varepsilon$ times the central value of $p_1$,
\begin{equation}
\phi_1^{\mathrm{cent}}=\varepsilon p_1^{\mathrm{cent}} \ . \label{p1centcond}
\end{equation}

Knowing $p_1$ as a function of $\rho$, we can calculate the scalar field $\phi$ by \eqref{phiepsexp} up to $\varepsilon^2$ order using \eqref{phi1eq} and \eqref{phi2eq}. Then $p_3$ yields $\phi$ to $\varepsilon^4$ order, using \eqref{phi3eq} and \eqref{phi4eq}. It is relatively easy to continue similarly to higher orders, by obtaining the relevant equations using an algebraic manipulation software for $p_{2n+1}$, and numerically integrating the resulting lengthy differential equations. The requirement \eqref{p1centcond} will yield the values of $\omega_{2n}$.

\subsection{Leading order behavior}

To leading order in the small-amplitude parameter the structure of oscillons is described by solutions of equation \eqref{p1eq}. We only have to consider localized solutions, in which case $p_1$ tends to zero exponentially at spatial infinity. We also assume that the solution is regular, i.e.~at the center of symmetry the derivative of $p_1$ is zero.

The $k=0$ Minkowski case has been studied in detail in \cite{FFHL2}. In that case regular localized solutions of \eqref{p1eq} exist only if $\omega_2<0$. Then by rescaling the radial coordinate $\rho$ one can set $\omega_2=-1$. At the same time, in order to keep the value of $\gamma$ unchanged, one has to rescale $p_1$ as well. These rescalings correspond to different choices of parametrizations of the various $\omega$ frequency states by $\varepsilon$. It can be shown that on Minkowski background it is possible to set all higher $\omega_n$ zero, and to make $\tilde\omega^2=1-\varepsilon^2$. Then it is possible to prove that localized solutions exist only if $\gamma>0$. Introducing $S=p_1\sqrt{\gamma}$, equation \eqref{p1eq} on a Minkowski background can be written as
\begin{equation}
S_{,\rho\rho}+\frac{D-1}{\rho}S_{,\rho}-S+S^3=0 \ . \label{seqmink}
\end{equation}
For $D=1$ spatial dimension the only localized solution is $S=\sqrt{2}\,\mathrm{sech}\,\rho$. For $D=2$ and $D=3$ there are various solutions enumerated by the number of zero crossings (nodes) of $S$. The only stable configuration corresponds to the nodeless solution. For $D>3$ there are no localized regular solutions of \eqref{seqmink}.

If $k=iH$ and $\kappa=ih$ are imaginary, then our equations describe a scalar field on a de Sitter background. In that case, at large distances the $h^2\rho^2p_1$ term dominates in \eqref{p1eq}, and all solutions have an oscillating tail at large distances. However, when $H$ is small enough, there still exist very long living oscillon configurations, which lose energy very slowly by scalar field radiation\cite{fggirs}.

On an anti-de Sitter background, when $k>0$, there are localized solutions even if $\omega_2\geq1$. Hence, instead of fixing $\omega_2$, in this case we use the rescaling freedom in $\rho$ to set
\begin{equation}
\kappa=1
\end{equation}
in \eqref{p1eq}. According to \eqref{kepmkappa}, this means that our small parameter $\varepsilon$ is related to the cosmological parameter $k$ by
\begin{equation}
k=\varepsilon^2m \ .  \label{kep2m0}
\end{equation}
One may consider this as the definition of $\varepsilon$ in the AdS case. Since we expand $\phi$ according to \eqref{phiepsexp}, our expansion is a small $k$ and small amplitude expansion at the same time, with one common small parameter.

In the AdS case there are localized solutions even if $\gamma\leq0$. Hence there are three different cases to deal with according to the signature of $\gamma$. The parameter $\gamma$ describes the behavior of the scalar potential $U(\phi)$ near its minimum. If $\gamma>0$, as in the case of the $\phi^4$ potential \eqref{phi4pot}, the potential well is less steep than that of the Klein-Gordon potential $U(\phi)=m^2\phi^2/2$. In such potential the oscillation frequency of a point mass would be smaller than in a same mass harmonic potential. We will see that a positive $\gamma$ will act as an effective attractive force on oscillons, making their size more compact, while a negative $\gamma$ will have a repulsive effect.

\begin{figure}[!hbtp]
\includegraphics[width=10cm]{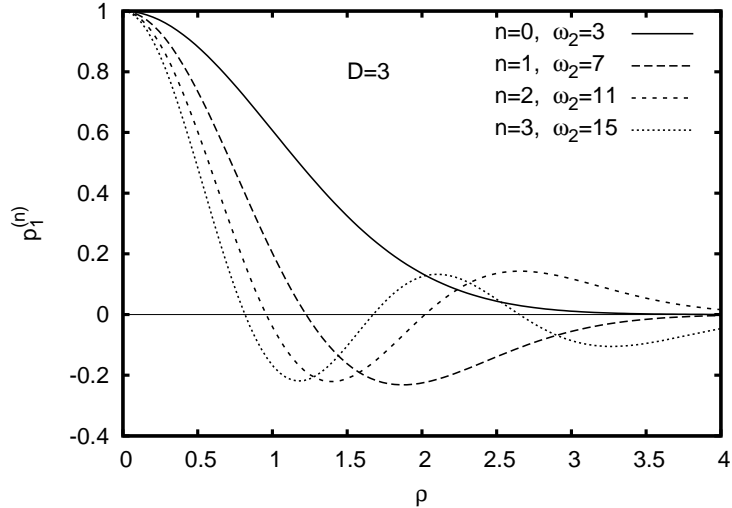}
\caption{Localized solutions for $p_1$ in the $\gamma=0$ case. \label{figseqlin}}
\end{figure}

If $\gamma=0$ the leading order behavior is the same as that of the Klein-Gordon field. In this case \eqref{p1eq} becomes linear,
\begin{equation}
p_{1,\rho\rho}+\frac{D-1}{\rho}p_{1,\rho}+\left(\omega_2-\rho^2\right)p_1=0 \ .
\end{equation}
This has regular localized solution only if
\begin{equation}
\omega_2=D+4n \ , \label{om2dn}
\end{equation}
where $n$ is a non-negative integer, and the solution is
\begin{equation}
p_1^{(n)}=\frac{n!}{(D/2)_n}
\exp\left(-\frac{\rho^2}{2}\right)L_{n}^{D/2-1}(\rho^2) \ , \label{laguerre}
\end{equation}
where $L$ denotes the generalized Laguerre polynomial. The normalization factor, containing the Pochhammer symbol $(D/2)_n$, is included in order to scale the central value to one. The solution $p_1^{(n)}$ has $n$ nodes. The first few solutions are
\begin{align}
 p_1^{(0)}&=\exp\left(-\frac{\rho^2}{2}\right)  \label{breatherp0} \ , \\
 p_1^{(1)}&=\left(1-\frac{2\rho^2}{D}\right)\exp\left(-\frac{\rho^2}{2}\right) \ , \\
 p_1^{(2)}&=\left(1-\frac{4\rho^2}{D}+\frac{4\rho^4}{D(D+2)}\right)
  \exp\left(-\frac{\rho^2}{2}\right) \ , \\
 p_1^{(3)}&=\left(1-\frac{6\rho^2}{D}+\frac{12\rho^4}{D(D+2)}-\frac{8\rho^6}{D(D+2)(D+4)}\right)\exp\left(-\frac{\rho^2}{2}\right) .
\end{align}
For $D=3$ spatial dimensions the functions are plotted on Fig.~\ref{figseqlin}.
Since to leading order the scalar field is $\phi=\varepsilon \phi^{(1)}=\varepsilon p_1\cos\tau$, we have essentially re-calculated the Klein-Gordon breather solutions of section \ref{kleingordon}, with the assumption that the cosmological parameter $k$ is small. Expression \eqref{laguerre} can also be obtained by taking the small $k$ limit of \eqref{kgbreather}, using $\cos x=1/\sqrt{(1+k^2r^2)}$, substituting $k$ from \eqref{kep2m0}, $r$ from \eqref{rhortrans}, and taking the small $\varepsilon$ limit. The necessary condition \eqref{om2dn} on the frequency shift parameter $\omega_2$ can be also obtained by taking the small $k$ limit of \eqref{lambdapcond}.

\begin{figure}[!htbp]
\includegraphics[width=10cm]{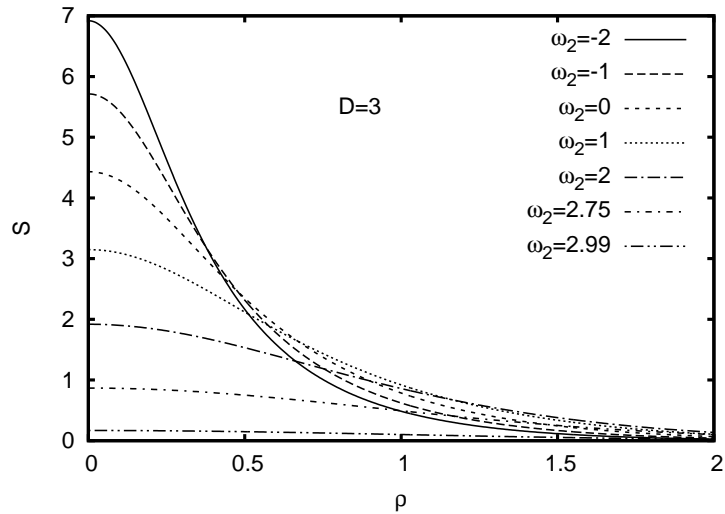}
\caption{Localized nodeless solutions for $S$ in the $\gamma>0$ case. \label{figseqlp}}
\end{figure}

If $\gamma>0$, then introducing $S=p_1\sqrt{\gamma}$, equation \eqref{p1eq} can be written as
\begin{equation}
S_{,\rho\rho}+\frac{D-1}{\rho}S_{,\rho}+(\omega_2-\rho^2)S+S^3=0 \ . \label{seqlapos}
\end{equation}
This equation has regular localized nodeless solution if and only if $\omega_2<D$, and for given $\omega_2$ the solution is unique. Localized solutions for various $\omega_2$ in $D=3$ dimensions are plotted on Fig.~\ref{figseqlp}.
The $\gamma>0$ case corresponds to an attractive scalar potential. Higher amplitude solutions are more localized because of the attraction provided by the potential. In this case the potential is less steep near its minimum than that of the same mass Klein-Gordon potential. In such potential a point mass would oscillate with a lower frequency than that of the harmonic oscillator.

\begin{figure}[!htbp]
\includegraphics[width=10cm]{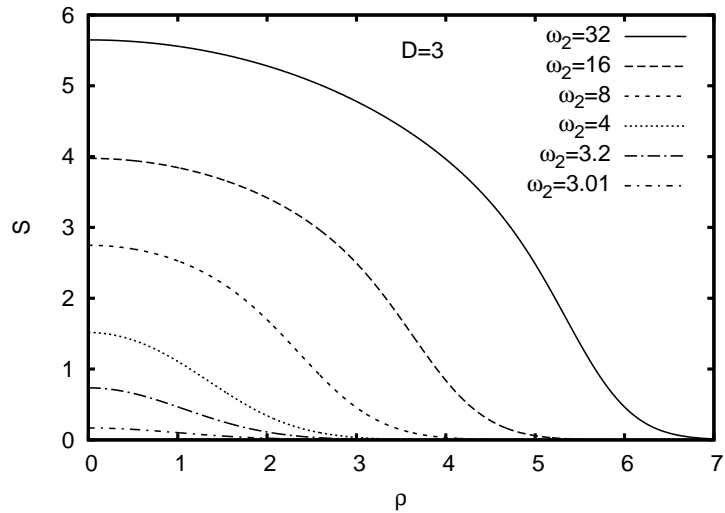}
\caption{Localized nodeless solutions for $S$ in the $\gamma<0$ case. \label{figseqlm}}
\end{figure}

\begin{figure}[!htbp]
\includegraphics[width=10cm]{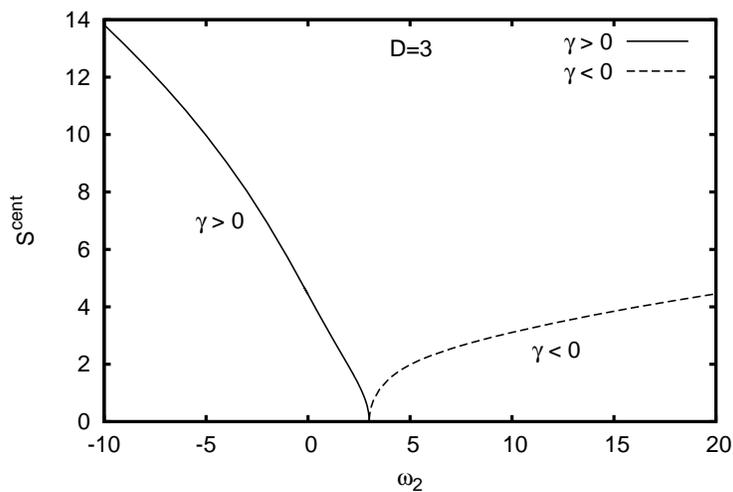}
\caption{Value of $S$ at $\rho=0$ as a function of $\omega_2$. The amplitude tends to zero at $\omega_2=3$. \label{scamplomfig}}
\end{figure}

The $\gamma<0$ case corresponds to a repulsive potential. In this case, introducing $S=p_1\sqrt{-\gamma}$, equation \eqref{p1eq} can be written as
\begin{equation}
S_{,\rho\rho}+\frac{D-1}{\rho}S_{,\rho}+(\omega_2-\rho^2)S-S^3=0 \ . \label{seqlaneg}
\end{equation}
This equation has regular localized nodeless solution if and only if $\omega_2>D$, and the solution is unique. The solutions for various $\omega_2$ in $D=3$ dimensions are plotted on Fig.~\ref{figseqlm}.
In this case, the larger the amplitude is, the bigger the size of the oscillon becomes. The shape of small amplitude solutions, when $\omega_2$ becomes close to $D$, approaches that of the Klein-Gordon breather \eqref{breatherp0} in both the $\gamma<0$ and $\gamma>0$ cases. On Fig.~\ref{scamplomfig} we show the value $S$ at the symmetry center as a function of $\omega_2$.

\subsection{Comparison to numerical results}

\begin{figure}[!htbp]
\includegraphics[width=10cm]{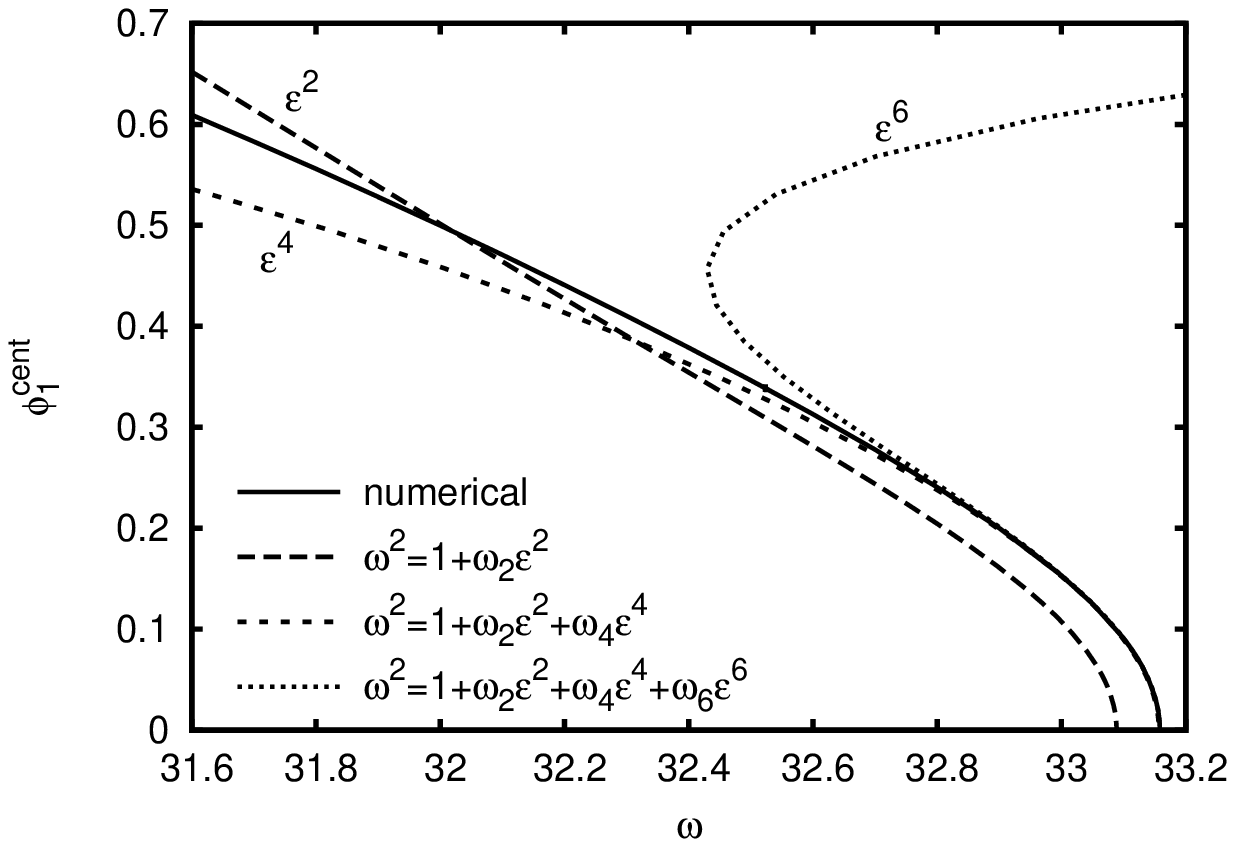}
\caption{Frequency dependence of the central value of the first Fourier mode for $k^2=0.001$. The solid line shows the precise numerical value obtained by the spectral code. The dashed lines show various order approximations calculated by the small $k$ expansion. \label{ordsphi1fig}}
\end{figure}

In order to get information about the validity region of the small $k$ expansion procedure, we compare the results to the $k^2=0.001$ periodic states that were obtained in Sec.~\ref{selfinter} by a spectral numerical code for the $\phi^4$ potential \eqref{phi4pot} in $D=3$ spatial dimensions. For this potential $m=1$, $\tilde g_2=-3/2$, $\tilde g_3=1/2$, and $\gamma=3/2$. In this case $\varepsilon=\sqrt{k}\approx0.1778$. The relation between the central value of $S=p_1\sqrt{3/2}$ and the frequency shift parameter $\omega_2$ has been already given on Fig.~\ref{scamplomfig}. The dependence of $\omega_4$ and $\omega_6$ on $p_1^\mathrm{cent}$ can be calculated by numerically finding the localized centrally vanishing solutions of \eqref{p3eq} for $p_3$, and of a corresponding lengthier equation for $p_5$. Then the dependence of the frequency $\omega=m\tilde\omega$ on the amplitude can be calculated using the expansion \eqref{omegatexp}. The comparison of the numerical and small $k$ expansion results to various orders can be seen on Fig.~\ref{ordsphi1fig}.
Contrary to the small-amplitude expansion performed in Sec.~\ref{smallamplsec}, for the small $k$ expansion used in this section the frequency in the small amplitude limit is given by an expansion only, so the higher the expansion order is the more precise the frequency becomes where the lines intersect the $\phi_1^{\mathrm{cent}}=0$ axis. The $\varepsilon^4$ order approximation gives a quite good approximation in the frequency interval shown on the plot. The $\varepsilon^6$ order approximation gives even better result for $\omega/k>32.9$, but for larger amplitude states its error grows quickly. This indicates that the small $k$ expansion is as asymptotic one, which is the case in general for the small-amplitude expansion of oscillons on flat background.

For small values of the cosmological constant it is much easier to apply the expansion procedure of this section than the small-amplitude expansion procedure of Sec.~\ref{smallamplsec}. In this case only one new differential equation appears at each odd $\varepsilon$ order, and the equations are simpler and easier to generate by an algebraic manipulation program. Another advantage is that if the dependence of the $\omega_i$ coefficients on the central amplitude is calculated once, then the result can be applied to arbitrary $k$, assuming that $k$ is small enough. For the small-amplitude expansion of Sec.~\ref{smallamplsec} the numerical integration of the differential equations has to be performed for each $k$ separately.

\section{Conclusions}
We construct by numerical and perturbative methods spatially localized, time-periodic, spherically symmetric breather-type solutions of scalar field theories with various self-interacting potentials on Anti-de Sitter (AdS) spacetime in $D$ spatial dimensions. The perturbative construction is a small amplitude expansion around
the scalar eigenfunction of the linear Klein-Gordon equation on AdS spacetime. 
The AdS breathers we construct form one-parameter families parametrized by their amplitude, $\varepsilon$, and by their frequency, $\omega=\omega(\varepsilon)$. High precision spectral methods allow us to obtain
AdS breathers by solving the nonlinear equations obtained by a Fourier decomposition in time directly and compare
them to the perturbative construction. It is found that the perturbation theory works very well, not only for small
but even for relatively large ($\varepsilon\lesssim 0.25$) amplitudes. 
Importantly the AdS breathers appear to be generically stable under time evolution.
A detailed numerical study of large amplitude configurations in $D=3$ reveals the appearance of resonances for
certain special values of the frequency and of the cosmological constant. These resonances merit further numerical and mathematical investigations.
We also implement a suitable small-amplitude expansion for small values of the cosmological constant,
suitable to study the rather singular Minkowskian limit.

\begin{acknowledgments}
This research has been supported by OTKA Grant No. K 101709.
\end{acknowledgments}


\begin{thebibliography}{99}


\bibitem{bizon-11}
P. Bizo\'{n} and A. Rostworowski, {\em Phys. Rev. Lett.}, {\bf 107}, 031102 (2011).

\bibitem{dias-12}
O. J. C. Dias, G. Horowitz and J. E. Santos, {\em Class. Quantum Grav.} {\bf 29}, 194002 (2012).

\bibitem{wheeler-55}
J. A. Wheeler, {\em Phys. Rev.} {\bf 97}, 511 (1955).

\bibitem{brill-64}
D. R. Brill,  and  J. B. Hartle, {\em Phys. Rev.} {\bf 135}, B271 (1964).

\bibitem{dias-12b}
O. J. C. Dias, G. Horowitz, D. Marolf and J. E. Santos, {\em Class. and Quantum Grav.} {\bf 29}, 235019 (2012).

\bibitem{maliborski}
M. Maliborski and A. Rostworowski,  {\em Phys. Rev. Lett.} {\bf 111}, 051102 (2013).

\bibitem{AdStimedep}
G. Fodor, P. Forg\'acs and P. Grandcl\'ement, Scalar fields on anti-de Sitter background, to appear in the proceedings of the conference {\em Relativity and Gravitation, 100 Years after Einstein in Prague}. The slides of the talk can be downloaded from ae100prg.mff.cuni.cz/presentations/Fodor\_Gyula.pdf

\bibitem{avis} S. J. Avis, C. J. Isham and D. Storey, \emph{Phys. Rev. D} {\bf 18}, 3565 (1978).

\bibitem{isibashiwald} A. Ishibashi and R. M. Wald, \emph{Class. Quantum Grav.} {\bf 21}, 2981 (2004).

\bibitem{BogMak2} I. L. Bogolyubskii and V. G. Makhan'kov,
{\em JETP Letters} {\bf 25}, 107 (1977).

\bibitem{Gleiser} M. Gleiser,
{\em Phys. Rev. D} {\bf 49}, 2978 (1994).

\bibitem{CopelGM95} E. J. Copeland, M. Gleiser and H.-R. M\"uller,
 {\em Phys. Rev. D} {\bf 52}, 1920 (1995).

\bibitem{Honda} E. P. Honda and M. W. Choptuik,
{\em Phys. Rev. D} {\bf 65}, 084037 (2002).

\bibitem{oscillons}
G. Fodor, P. Forg\'acs, P. Grandcl\'ement and I. R\'acz, {\em Phys. Rev. D} {\bf 74}, 124003 (2006).

\bibitem{FFHL2} G. Fodor, P. Forg\'acs, Z. Horv\'ath, \'A.  Luk\'acs,
{\em Phys. Rev. D} {\bf 78}, 025003 (2008).

\bibitem{SK} H. Segur and M. D. Kruskal, \emph{Phys. Rev. Lett.}
  \textbf{58}, 747 (1987).

\bibitem{FFHM} G. Fodor, P. Forg\'acs, Z. Horv\'ath and M. Mezei, {\em
    Phys. Rev. D} \textbf{79}, 065002 (2009).

\bibitem{moredim} G. Fodor, P. Forg\'acs, Z. Horv\'ath and M. Mezei,
  \emph{Phys. Lett. B} \textbf{674}, 319 (2009).

\bibitem{Seidel1} E. Seidel and W-M. Suen, \emph{Phys. Rev. Lett.}
  \textbf{66}, 1659 (1991).

\bibitem{Seidel2} E. Seidel and W-M. Suen, \emph{Phys. Rev. Lett.}
  \textbf{72}, 2516 (1994).

\bibitem{page04} D. N. Page, {\em Phys. Rev. D} \textbf{70}, 023002 (2004).

\bibitem{oscillaton10} G. Fodor, P. Forg\'acs and M. Mezei,
{\em Phys. Rev. D} {\bf 82}, 044043 (2010).

\bibitem{oscillatons}
P. Grandcl\'ement, G. Fodor, and P. Forg\'acs, {\em Phys. Rev. D} {\bf 84}, 065037 (2011).

\bibitem{GenMastroProc}
G. Gentile, V. Mastropietro, M. Procesi, {\em Comm. Math. Phys.} {\bf 256}, 437 (2005).

\bibitem{dlmf}
Digital Library of Mathematical Functions, National Institute of Standards and Technology, Chapter 15, http://dlmf.nist.gov/

\bibitem{balasub} V. Balasubramanian, P. Kraus and A. Lawrence, \emph{Phys. Rev. D} {\bf 59}, 046003 (1999).

\bibitem{kadathpaper}
P. Grandcl\'ement, \emph{J. Comput. Phys.} {\bf 229}, 3334 (2010).

\bibitem{kadath}
{\tt KADATH} web-site : {\em http://luth.obspm.fr/$\sim$luthier/grandclement/kadath.html}

\bibitem{living} P. Grandcl\'ement and J. Novak,
{\em Living Reviews in Relativity}, lrr-2009-1 (2009).

\bibitem{FR} G. Fodor and I. R\'acz, {\em Phys. Rev. Lett.} {\bf 92}, 151801
(2004).

\bibitem{FR2} G. Fodor and I. R\'acz, {\em Phys. Rev. D} {\bf 77}, 025019
(2008).

\bibitem{dilaton} G. Fodor, P. Forg\'acs, Z. Horv\'ath and M. Mezei,
  JHEP08(2009)106 (2009).

\bibitem{dashen} R. F. Dashen, B. Hasslacher and A. Neveu,
{\em Phys. Rev. D} {\bf 11}, 3424 (1975).

\bibitem{Kichenassamy} S. Kichenassamy,
{\em Comm. Pur. Appl. Math.} {\bf 44}, 789 (1991).

\bibitem{fggirs} E. Farhi, N. Graham, A. H. Guth, N. Iqbal,
  R. R. Rosales and N. Stamatopoulos,
{\em Phys. Rev. D} {\bf 77}, 085019 (2008).



\end{thebibliography}
\end{document}